\documentclass[12pt,preprint]{aastex}
\usepackage{psfig}
\begin{document}
\def\dd{\partial}
\def\lta{\mathrel{\rlap{\lower 3pt\hbox{$\mathchar"218$}}
    \raise 2.0pt\hbox{$<$}}}

\shorttitle{Gap Formation in Turbulent Disks}
\shortauthors{Winters, Balbus, \& Hawley}

\title{Gap Formation by Planets in Turbulent Protostellar Disks}

\author{Wayne F. Winters\altaffilmark{1},
Steven A. Balbus\altaffilmark{1},
and John F. Hawley\altaffilmark{1}
}

\altaffiltext{1}{Dept. of Astronomy, University of Virginia, PO Box
3818, Charlottesville,\\ VA 22903, USA. jh8h@virginia.edu;
sb@virginia.edu}

\begin{abstract}

The processes of planet formation and migration depend intimately on
the interaction between planetesimals and the gaseous disks in which
they form.  The formation of gaps in the disk can severely limit the
mass of the planet and its migration toward the protostar.  We
investigate the process of gap formation through magnetohydrodynamic
simulations in which internal stress arises self-consistently from
turbulence generated by the magnetorotational instability.  The
simulations investigate three different planetary masses and two disk
temperatures to bracket the tidal (thermal) and viscous gap opening
conditions.  The results are in general qualitative agreement with
previous simulations of gap formation, but show significant
differences.  In the presence of MHD turbulence, the gaps produced are
shallower and asymmetrically wider than those produced with pure
hydrodynamics.  The rate of gap formation is also slowed, with
accretion occurring across the developing gap.  Viscous hydrodynamics
does not adequately describe the evolution, however, because planets
capable of producing gaps also may be capable of affecting the level
MHD turbulence in different regions of the disk.

\end{abstract}


\section{Introduction}

Understanding the process of planet formation in nascent solar systems
is a long-standing goal of astrophysical theory.  The traditional
picture is an orderly one in which planets slowly build up their mass
first by accreting rocky planetesimals, then (if sufficiently massive)
gas from the surrounding protostellar disk.  The recent multiple
detections of extrasolar planets close to the central star, a
configuration once thought to be highly improbable, pose a stern
challenge to this relatively simple picture of planet formation.  In
the wake of these discoveries, contemporary theories emphasize the
importance of the dynamical interaction between a developing planet and
the ambient gaseous disk.

A possible explanation for the presence of giant planets near their
central star is that they have migrated inwards after having formed
farther out in the disk.  This requires angular momentum to be removed
from the planet by an imbalance in the positive and negative torques
exerted upon it \citep{gt80, w86, wd97}.  Two mechanisms have been
identified, known as Type I and Type II migration.  Type I migration
is essentially a linear process in which the planet excites trailing
density waves at the Lindblad resonances.  The waves carry angular
momentum outward through the disk.  There are torques of opposite signs
exerted at both the inner and outer resonances, but for a given $m$
(azimuthal wavenumber), the outer resonance lies closer to the planet and
the coupling is stronger \citep{wd97}.  Therefore, if the interaction is
dominated by the Lindblad torques, then more angular momentum is lost
than acquired by the planet, which would then drift inward.  The role
of corotation torques, however, is still under investigation, and needs
to be understood if Type I migration is to be put on a firm theoretical
footing \citep{ttw02}.  In Type II migration, nonlinear gravitational
interactions between the disk and the planet clear a gap at the orbital
radius of the planet \citep{lp79, LP93, gt80}.  Disk material then no
longer accretes across the planet's orbital radius, and the planet is
carried inward with the disk itself on an accretion time scale.  The rate
of migration thus depends on whether a gap forms or not, and whether
there is, in fact, an turbulent stress to drive accretion in the disk.
Unless for some reason disk material remains abundant in the immediate
vicinity of the planet, one important consequence of gap formation is
likely to be that planet formation will be severely limited.

Traditionally, turbulence has been modeled as a Stokesian viscosity
in accretion disk studies, and the criteria for a planet to open
a gap are not the same in viscous and inviscid models.  \cite{LP93}
and \cite{BCLNP99} discuss the conditions pertinent to each class of
model.  In the case of an inviscid accretion disk, the competing forces
are the local fluid pressure, which resists the creation of a gap,
and the planet's gravitational tidal force, which promotes gap formation.
The tidal criterion for gap formation is that a gap can form if the
planet's Roche (or Hill) radius $R_R$ exceeds
the pressure scale height $H$, which we define as the ratio of
isothermal sound $c_s$ to local orbital frequency $\Omega$.  
For a planet of mass $M_p$
orbiting a central star of mass $M_*$ at radius $R_p$, the Roche limit
is $R_R = (M_p/3M_*)^{1/3}R_p$.  Thus a gap forms if
\begin{equation}\label{tidal}
q\equiv \frac{M_{p}}{M_{*}} \gtrsim 3
\left(\frac{H}{R_p}\right)^{3}.
\end{equation}
Denoting the right hand side of equation (\ref{tidal}) as $q_t$, 
the tidal criterion is simply $q\gtrsim q_t$.

In viscous disk models, there is another condition that must be
satisfied before a gap can open.  The is set by the requirement that
the rate at which planetary torques clear the gap must exceed the rate
at which viscous torques replenish it.  The condition is \citep{LP93}
\begin{equation}
q \gtrsim q_v \equiv
40 \alpha \left(\frac{H}{R_p}\right)^{2},
\label{eq:viscous}
\end{equation}
\noindent 
where $\alpha$ is defined as the ratio of the 
viscosity to $c_s H$.
This is essentially the \cite{SS73} parameterization

These gap conditions have been extensively investigated via
two-dimensional numerical simulations of $\alpha$ disks \citep{BCLNP99,
kl99, LSA99}.  An important limitation of this approach is that the most
interesting consequences of the planet-disk interaction depend upon the
nature of the stress, and viscosity is not synonymous with turbulence.
The origin of the anomalous disk stress and turbulent transport is no
longer unapproachable.  If $\alpha$ represents a stress arising from
internal disk turbulence, then the source of this turbulence is magnetic
\citep{BH91}.  As discussed by \cite{BH98} purely hydrodynamic sources
of turbulence are problematic at best, and none has been observed and
sustained in a fully three-dimensional simulation of a Keplerian disk.
The problem is illustrated by studies of convective turbulence: when
this form of turbulence is imposed, it drives inward, not outward,
angular momentum transport \citep{sb96}.  This is no fluke of the
peculiar properties of convection, it is what happens quite generally
if incompressible turbulence is driven in a rotationally stable disk.
If nonlinear instabilities were a possibility, they would be present in
convective and shearing sheet simulations \citep{HBW99}.  The shearing
sheet is scale-free and any instabilities would accordingly not require
extremely large Reynolds numbers to appear.  (There is nothing
special about the behavior of very small scale disturbances that
would not also be seen at larger scales.)  This is why \cite{HBW99} easily
found nonlinear planar Couette flow instabilities.  Their absence
from any numerical simulation of a Keplerian disk casts doubt on their
existence. Trailing density waves transport angular momentum outwards of
course, but their excitation and maintenance must be explained, and this
is most readily done only if the disk itself is strongly self-gravitating.
Even in this case, the transport properties are generally nonlocal,
and not well-suited to an $\alpha$ formalism description \citep{BP99}.
The nature of, and difficulties with, an $\alpha$ disk associated
with direct planetary torques have recently been discussed by
\cite{gr01} (see also \cite{lp96}).  

In this paper we perform numerical experiments of planets in protostellar
disks in which the internal turbulence and its stress arise self-consistently
from the magnetorotational instability.  These simulations examine
how turbulence affects the formation of gaps in the disk, and, equally
importantly, how the planet affects the turbulence.  In this first study
we will assume ideal MHD, and not address the issue of the ionization
state of the disk, or the degree of magnetically coupling in the gas.
These issues are far from resolved \citep{ftb02}.  It is of interest to
examine the ``simplest'' case first.

The plan of this paper is as follows:  in \S2 we describe the numerical
model used for the simulations.  In \S3 we present several two
dimensional hydrodynamic simulations.  These models serve as code
tests, they provide a comparison with previous numerical work, and they
are the controls for the MHD simulations which follow in \S4.  We
present our conclusions in \S5.

\section{Numerical Model}

The governing equations for the evolution of a magnetized
gaseous disk in the ideal MHD limit are the equation of continuity,
\begin{eqnarray}
\frac{\partial \rho}{\partial t} + \nabla \cdot \left( \rho {\mathbf v}
\right)
&=& 0,
\end{eqnarray}
the equation of motion,
\begin{eqnarray}
\rho \frac{\partial {\mathbf v}}{\partial t}
+(\rho {\mathbf v} \cdot \nabla){{\mathbf v}}
&=& - \nabla \left(P + \frac{B^{2}}{8 \pi}\right)
- \rho \nabla \Phi
+ \left(\frac{{\mathbf B}}{4 \pi} \cdot \nabla \right){\mathbf B},
\end{eqnarray}
and the induction equation
\begin{eqnarray}
\frac{\partial {\mathbf B}}{\partial t} &=&
\nabla \times \left({\mathbf v}\times {\mathbf B}\right).
\end{eqnarray}
In these equations $\rho$ is the disk mass density,
${\mathbf v}$ is the fluid velocity, $P$ is the pressure, 
${\mathbf B}$ is the magnetic field vector, and $\Phi$
is the gravitational potential.  
These equations are solved in a
cylindrical coordinate system, $(R,\phi,z)$,
in the ``cylindrical disk'' limit, in which
vertical gravitational forces are ignored.  
Time and length units are
fixed by setting $GM_* = 1$, which makes the Keplerian angular velocity,
$\Omega \left[ = (GM_*/R^{3})^{1/2}\right] = 1 $ at $R = 1$.

The equation of state is ``locally isothermal,'' with a prescribed
temperature profile, $T(R)\propto R^{-1}$.  This is the same
equation of state used in many of the previous planet-disk 
simulations, e.g., \cite{kl99, BCLNP99, NPMK00}.
Specifically,
\begin{equation}
P = \rho c_s^2 = \rho \left(\frac{H}{R}\right)^{2} \frac{GM_*}{R},
\end{equation}  
where ${H}/{R}$ is a fixed input parameter of a given simulation.  
In this model, the Mach number of the flow
($v_{\phi}/c_s$) is constant with radius.  The initial disk is given a
density profile $\rho \propto R$ with $\rho = 1$ at $R=1$.  This ensures
that the initial pressure gradient is zero.  The equations that are
explicitly integrated
also contain an artificial viscosity of the form described by
\cite{SN92a}.  This is nonzero only in regions of local compression
and acts only on the momentum, since the temperature at any given radius
is fixed.  Its principal function is to spread shock transitions
over a few grid zones.

The computational grid consists of a cylindrical annulus centered on the
central star, with the radial coordinate running between $0.25< R< 3.75$.
The use of an annulus avoids the singularity associated with $R=0$.
The angle $\phi$ runs over the full range from 0 to $2\pi$.  The vertical
grid size, $L_z$, is set equal to twice the scale height $H$ at the
radius of the planet, $R_p$, i.e., for $H/R=0.05$, $L_z=0.2$.  The three
dimensional grid resolution is $128\times 128\times 32$ in $(R,\phi,z)$.
We work in the ideal MHD limit of a single, infinitely conducting medium.

The gravitational potential
for the central star is $\Phi = -GM_*/R$.  
The planet is assumed to remain on a fixed
circular orbit with a fixed orbital velocity at radius $R_p = 2$.  
The planet's orbital frequency is
\begin{equation}
\Omega_p^{2} = \frac{G}{R_{p}^{3}}(M_{*}+M_{p}), \nonumber
\end{equation}
and the planet's angular location $\phi_p$, is a
function of time,
\begin{equation}
\phi_{p} =\phi_{0}+\Omega_p t. \nonumber
\end{equation}
The gravitational potential of the planet is that of a softened 
point particle, specifically,
\begin{equation}
\Phi_{p} = -GM_{p}/(d^{2}+b^{2})^{1/2},
\end{equation}
where $d$ is the distance from the planet to a
location within the grid, $d^{2} = R^{2}+R_{p}^{2}-2RR_{p}\cos(\phi-
\phi_{p})$, and $b$ is the softening
parameter set to be a fraction of the size of the planet's Roche radius.
In the simulations below, $b$ was set to be equal to one-fifth of the
size of the Roche radius for a planet with $q=0.001$,  corresponding to
$b=0.0277$.  The radial and angular gravitational force per unit volume
from a planet at a point $(R,\phi)$ are
\begin{eqnarray}
f_{R} &=&- \rho GM_{p}
\left[\frac{(R - R_{p}\cos(\phi-\phi_{p}))}
{(R^{2}+R_{p}^{2}-2RR_{p}\cos(\phi-\phi_{p})+b^{2})^{\frac{3}{2}}}
- \frac{\cos(\phi-\phi_{p})}{R_{p}^{2}}\right]\\
f_{\phi} &=& - \rho GM_{p}
\left[\frac{(R_{p}\sin(\phi-\phi_{p}))}
{(R^{2}+R_{p}^{2}-2RR_{p}\cos(\phi-\phi_{p})+b^{2})^{\frac{3}{2}}}
+ \frac{\sin(\phi-\phi_{p})}{R_{p}^{2}}\right].
\end{eqnarray}
Since the origin of the coordinate system is non-inertial, the potential
includes the indirect term (which accounts for the acceleration of the
computational frame).  Note that the planet is represented only as a
potential; there is no mass loss from the grid at its location, and we
do not attempt to model accretion onto the planet in detail.

We have run two types of simulation: inviscid hydrodynamic models in
two dimensions $(R,\phi)$, and three-dimensional turbulent MHD
simulations.  Because neither the MRI nor field amplification can
operate without vertical structure, simulations of MRI-induced
turbulence must be three-dimensional.  However, some simplification is
still required to make the problem tractable.  The cylindrical
disk approximation makes it practical to compute disks that have small
$H/R$ values without the need for an unduly large number of grid
zones.  This cylindrical limit was previously employed by \cite{jh01}
in simulations of MHD turbulence in cold Keplerian disks.

For the MHD simulations the initial magnetic field strength is set by
the input parameter, 
\begin{equation}
\beta = {8 \pi P}/{B^{2}},
\end{equation}
which specifies the ratio of the gas pressure to the magnetic pressure.
To minimize difficulties with the magnetic radial boundaries, the initial
magnetic field is confined to the region between $R=0.45$ and $R=3.55$.
The subsequent evolution does allow field to leave through the radial
boundaries, however.

The equations were solved using time-explicit Eulerian finite
differencing.  The numerical algorithm is that used by the ZEUS code
for hydrodynamics \citep{SN92a} and MHD \citep{SN92b,HS95}.  The
imposed boundary conditions are periodic along the azimuthal and
vertical directions.  The radial boundary conditions are somewhat more
complex and are set in ghost zones exterior to the inner and outer
radial limits of the disk.  The density $\rho$ in the ghost zones is
held fixed in time at its initial value.  The azimuthal velocity
$v_{\phi}$ in the ghost zones is held constant and equal to the
appropriate Keplerian value.  Both the $v_{r}$ and $v_{z}$ velocity
components in the ghost zones are set equal to the last physical values
computed at the adjacent boundary.  In the 3D MHD cases, the magnetic
ghost radial boundaries are that $B_{\phi}$ and $B_{z}$ vanish, and
$B_{r}$ is set so that it is divergence-free in the ghost zones.
Although straightforward to implement, these boundary conditions are at
least partially reflecting, and therefore less than ideal.

The simulations produce a significant amount of both spatial and
temporal data.  To facilitate the analysis, two-dimensional data were
averaged azimuthally, and three-dimensional quantities were averaged
both vertically and azimuthally.  The definition for the average of
some quantity $X$ is
\begin{equation}\label{avgeq}
\langle X\rangle = \frac{\int X Rd\phi dz}{\int R d\phi dz}.
\end{equation}
These averages were computed at regular time intervals for each radial
grid zone, and the results used to construct space-time diagrams that
characterize the overall evolution.

\section{Hydrodynamic Tests}

We begin with a series of purely hydrodynamic simulations to test the
numerical model and to establish a baseline for comparison with the
results from MHD simulations.  Two-dimensional $(R,\phi)$ hydrodynamic
simulations have been carried out by a number of other researchers,
e.g, \cite{BCLNP99, LSA99, kl99, NPMK00}.  \cite{kl3d} performed
three-dimensional simulations and found that their results largely
agree with those from two dimensional simulations so long as the scale
height of the disk is comparable to or less than the Roche radius
$R_{\rm R}$ of the planet, which is the tidal condition for gap
formation.  For $H > R_{\rm R}$ the accretion rate onto the planet can
differ between two and three dimensional studies.

The disk temperature profile is set by the selected scale height
ratio.  The standard value, ${H}/{R}=0.05$, corresponds to a constant
Mach number of 20 throughout the disk.  The initial components of the
velocity are $v_{r}=0$ and $v_{\phi}=({GM_*}/{R})^{1/2}$.  Three
different planet masses are used, with mass ratios $q = 2 \times
10^{-4}$, to $1 \times 10^{-3}$, and $5 \times 10^{-3}$.  Other
parameters for the simulations may be found in Table \ref{hydrotbl}.
The three main simulations are labeled HSP, HMP, HLP for Hydrodynamic
Small, Medium, and Large mass Planet respectively.  The ratio of $q$ to
the tidal limit $q_t$  (eqn.~\ref{tidal}) is given for each problem.
For $H/R=0.05$ the small mass planet has $q < q_t$, the medium mass is
exceeds $q_t$, and the large mass amply satisfies the tidal criterion.
For the HMPH(ot) simulation the temperature is increased sufficiently
that the medium mass planet no longer meets the tidal condition for gap
formation.  In the HSPC(old) simulation the temperature is reduced so
that the small mass planet can satisfy the tidal condition.  Thus the
ensemble of simulations bracket $q_t$ both through different planet
masses and different values of $H/R$.

\begin{table}[t]
\begin{center}
\caption{Hydrodynamic Simulations}
\vspace{12pt}
\begin{tabular}{|l|c|c|c|c|}
\hline
Label & $q$ & ${H}/{R}$ & $q/q_t$ & $\tau_{\mathrm{gap}}$\\ \hline \hline
HSP & $2\times 10^{-4}$ & 0.05 &$0.52$& 530 orbits\\
HMP & $1\times 10^{-3}$ & 0.05 &$2.6$& 90 orbits\\
HLP & $5\times 10^{-3}$ & 0.05 &$13.0$& 40 orbits\\
HMPH & $1\times 10^{-3}$ & 0.09 &$0.45$& 310 orbits\\
HSPC & $2\times 10^{-4}$ & 0.02 &$8.3$& 140 orbits\\
\hline
\end{tabular}
\label{hydrotbl}
\end{center}
\end{table}

\begin{figure}[t]
\begin{center}
See 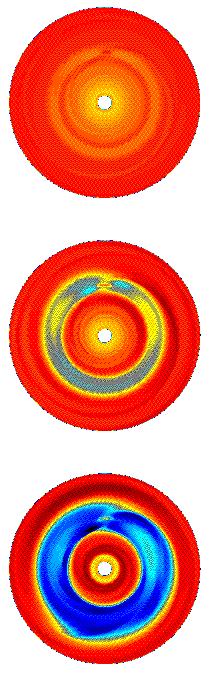
\caption{Log density in hydrodynamic simulations after 100 planet
orbits for (a) the small mass planet, $q=2\times 10^{-4}$, (b) the
medium mass planet, $q=1\times 10^{-3}$, and (c) the large mass
planet, $q=5\times 10^{-3}$.
The color map runs from blue to red in $\log(\rho)$ from
$\rho = 0.01$ to 1.0.
Note the density enhancements inside the gap at the trailing and
leading Lagrange points, L4 and L5.  }
\label{rd}
\end{center}
\end{figure}

Figure \ref{rd} shows the distribution of density after 100 planetary
orbits for the three simulations with the same $H/R$.  A qualitative
comparison of these plots illustrates how the width of the gap produced
changes with the mass ratio:  HSP fails to produce a gap, while HLP
produces an extensive one.  Although the small mass planet does not
create a gap, the density is nevertheless reduced within a region near
its orbit.  In the HSP run the reduced density region is spanned by
approximately 10 radial grid zones while the gaps in the HMP and HLP
runs are spanned by 30 and 45 radial grid zones,
respectively.  Also visible are the spiral density waves that the
planets generate.  The amplitude of these waves increases with
planetary mass.  Note too the density enhancements at the trailing and
leading Lagrange points, L4 and L5, particularly in the HMP run.

The density contours immediately surrounding the planet appear
elongated.  This is an artifact arising from the aspect ratio
of the $R$ and $\phi$ grid at the radius of the planet.  
The grid is uniformly spaced
in both $R$ and $\phi$ and (at the resolution used) $Rd\phi > dR$ at
the planet's orbital radius.  The zones are long thin rectangles, and
this results in an elongated planet shape.  To verify this, a test
calculation was run with the number of angular zones increased by a
factor of 3.5 giving $R_{p}d\phi = dR$ at the planet's radius.  With
this grid, the resulting density contours around planet are symmetric.
Ideally, one would prefer to have square zones at all radii.  This
would be particularly important in investigating the accretion of
material onto the planet.  The best approach might be a nested grid
around the planet \citep{LSA99, CPR00, DHK02}.  In the present
experiments, however, we are primarily concerned with the radial
structure in the disk, not the accretion onto the planet.  The grid we
use is a compromise between radial resolution and a practical amount of
computational time.

\begin{figure}[htp]
\begin{center}
\psfig{file=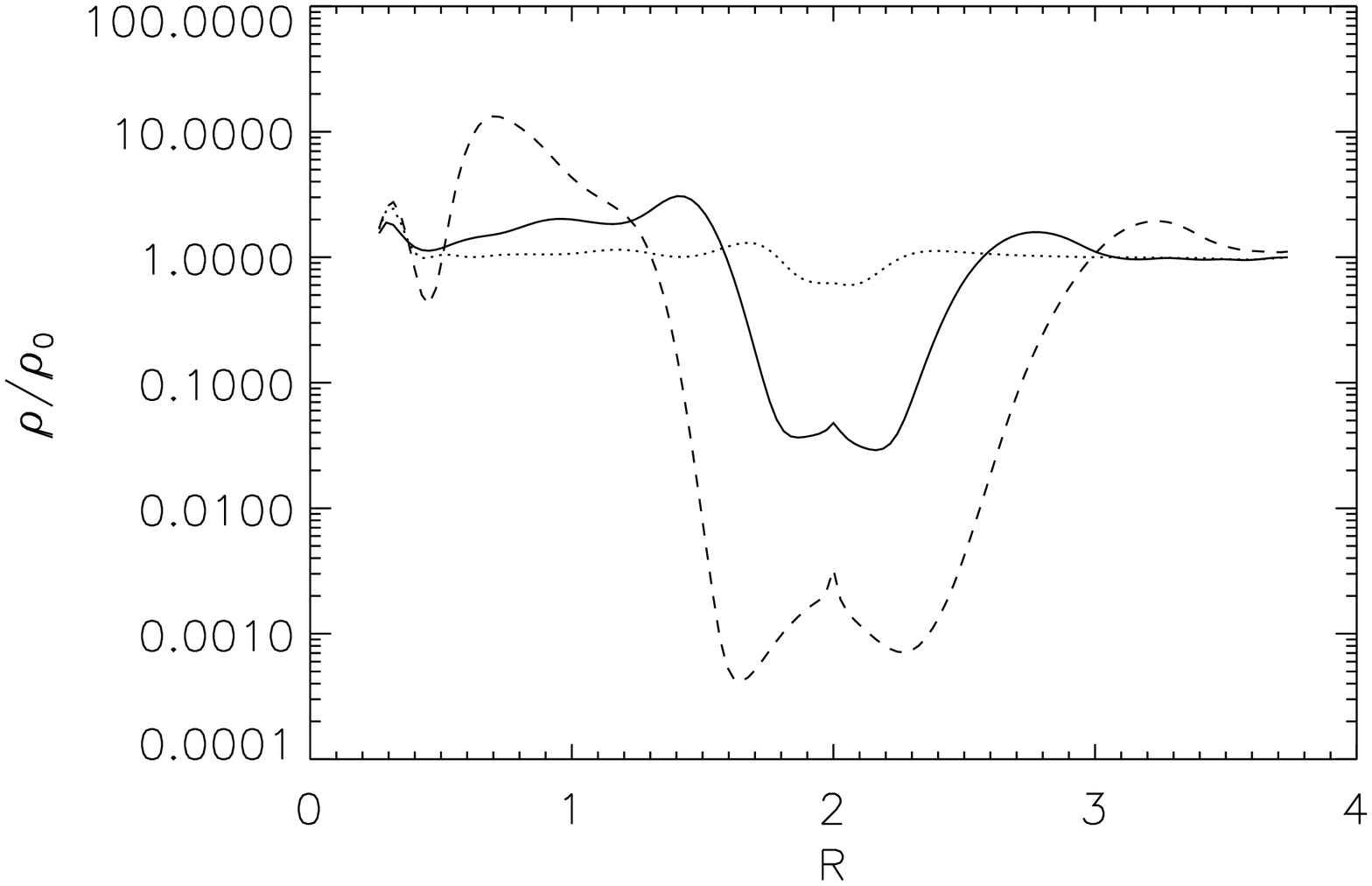,width=4.5in}
\psfig{file=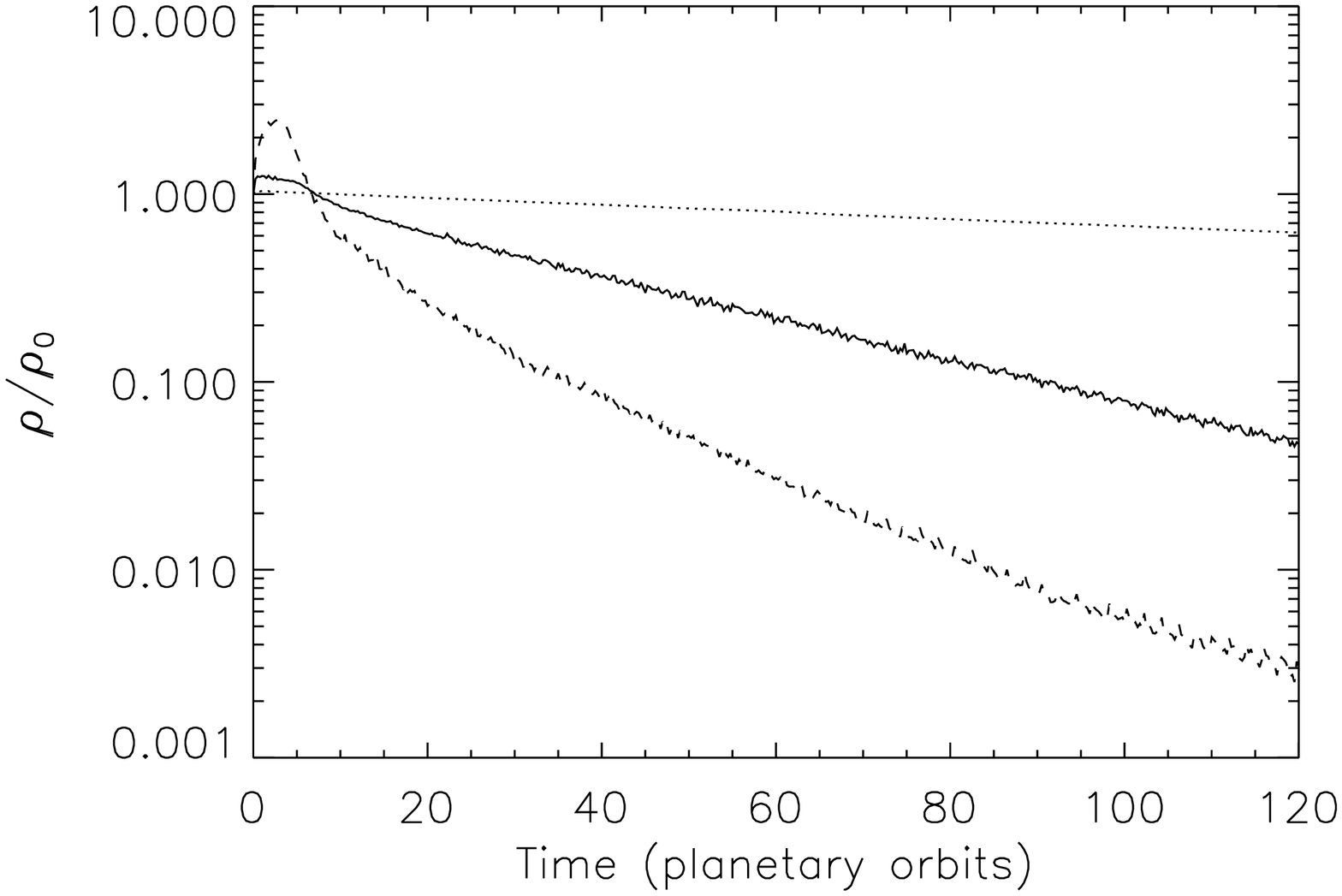,width=4.5in}
\caption{Time evolution of the gap density for hydrodynamic
(a) The azimuthally-averaged radial density profiles after 120ed
line).
planetary orbits for each run.  (b) The time
evolution of the azimuthally-averaged density at the planet's orbital
radius for each of the three simulations.}
\label{hyddentime}
\end{center}
\end{figure}

The top panel of Figure \ref{hyddentime} shows the radial density
profile (normalized by the initial profile) after 120 orbits for runs
HSP, HMP, and HLP. In both the middle and high mass cases a gap forms.
The bottom panel of Figure \ref{hyddentime} shows the time
history of the density evolution at the radial location of the planet.
The slope of the line is a measure of the rate of gap clearing;
not surprisingly this is
a function of planet mass.  The clearing rate may be characterized by
$\tau_{\mathrm{gap}}$, defined as the amount of time required for the
density at the planet's radius to decrease by a factor of 10.  The
value of $\tau_{\mathrm {gap}}$ is obtained by fitting the density
versus time at the planet's radius with a function of the form
${\rho}/{\rho_{o}} \sim 10^{-t/\tau_{\mathrm{gap}}}$, where $t$ is the
orbital time.  The value of $\tau_{\mathrm{gap}}$ is listed in Table
\ref{hydrotbl}.

\begin{figure}[htp]
\begin{center}
\psfig{file=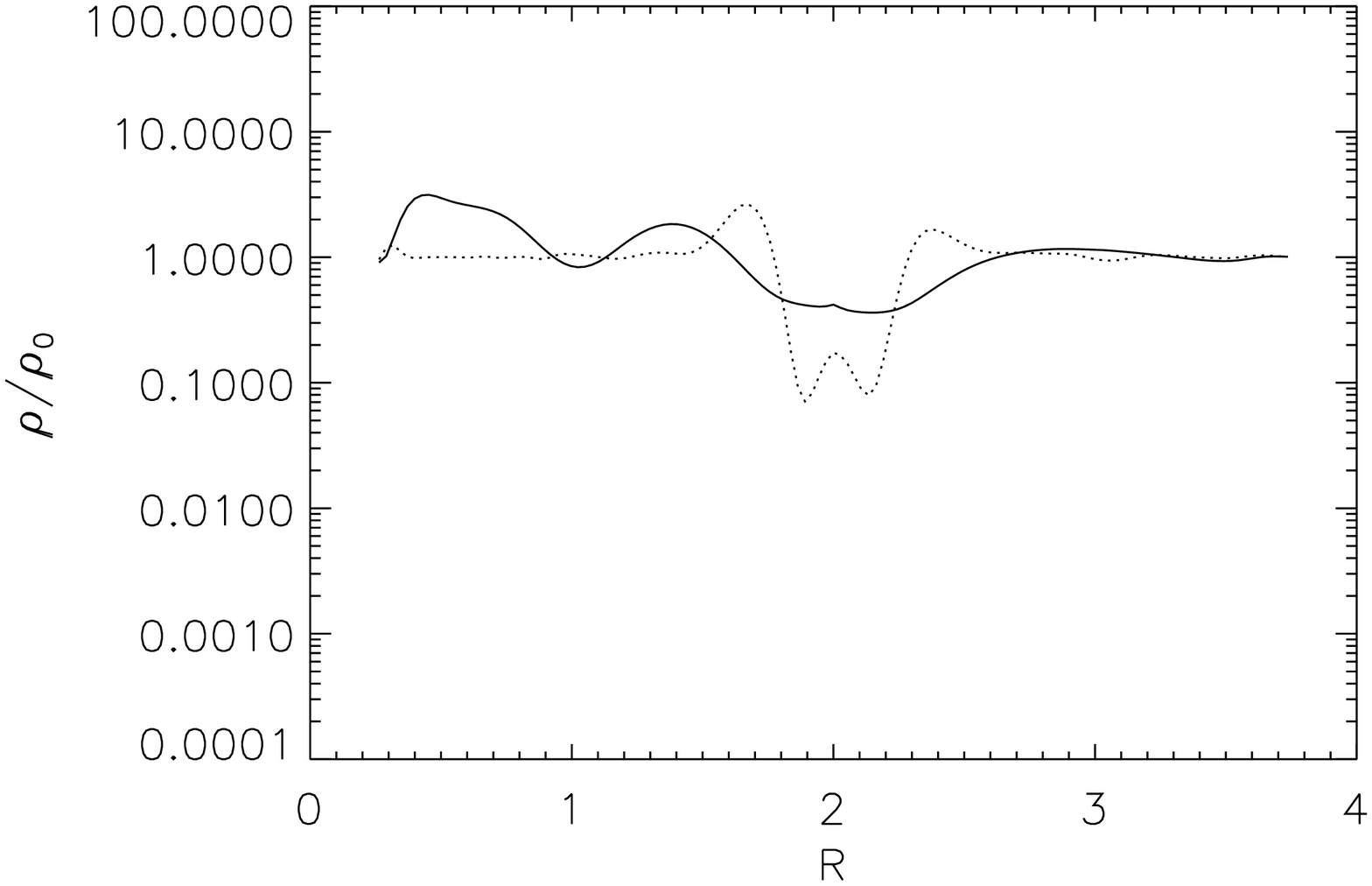,width=4.5in}
\psfig{file=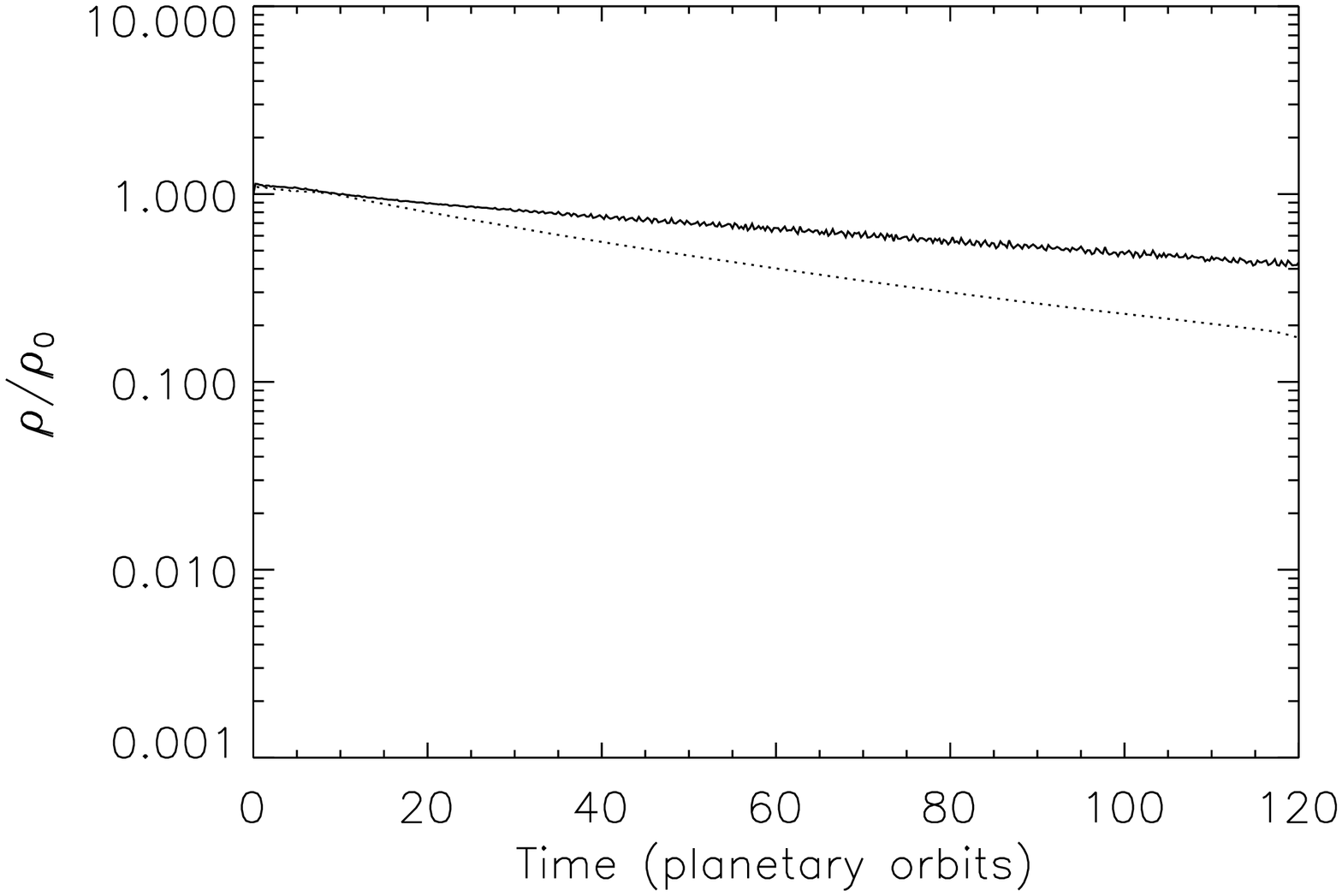,width=4.5in}
\caption{Time-evolution of the gap density for hydrodynamic
simulations HMPH
(solid line) and HSPC (dotted line).  (a) The
azimuthally-averaged radial density profiles after 120 planetary
orbits
for each run.  (b) The time evolution of the
azimuthally-averaged density at the planet's orbital radius for the
simulations.}
\label{hchyddentime}
\end{center}
\end{figure}

The three simulations above test the effect of varying the planet mass,
and two further hydrodynamical simulations test the tidal condition
with varying disk temperature.  In run HMPH the middle mass planet was
embedded in a disk that was too hot for it to produce a gap, and in run
HSPC the small mass planet was placed in a disk that was cool enough
to produce a gap.  Figure \ref{hchyddentime} is the density
profile and time-evolution for these two runs.  Comparing this data to
Figure \ref{hyddentime} one can see that cooling the disk increases the
rate at which the small mass planet clears its Roche radius, reducing
$\tau_{\rm gap}$ to 140 orbits.  Heating the disk greatly increases
$\tau_{\rm gap}$ for the middle mass planet, and eliminates the sharp
gap boundaries in the density distribution.

To summarize, these hydrodynamic simulations are in agreement with the
results of previous work and demonstrate the role of the tidal
condition in gap formation in stress-free disks.  The Roche radius
determines the width of the gap formed, and the temperature in the disk
determines whether that radius can be efficiently cleared by the
planet's gravity.  These disk simulations will serve as the baseline
comparisons for the series of disks with stress from MHD turbulence.

\section{MHD Turbulence and Gap Formation}

\subsection{MHD turbulent disks}

To establish a turbulent background flow, we begin with a planet-free
MHD simulation.  Previous work with cylindrical representations of MHD
turbulence \citep{jh01, sp02, np02a} provides some guidance for choosing a
suitable initial field configuration.  An initial vertical magnetic
field produces more powerful turbulence than does an initial toroidal
field, but also tends to produce a series of evacuated gaps in the
disk, even without the presence of a planet.  The extent to which this
is physical or an artifact of the cylindrical scheme is not yet
well-understood.  We have therefore elected to work primarily with
toroidal fields, though this is likely to underestimate the strength of
the turbulence actually present in a magnetized disk.

Our fiducial simulation starts with a uniform toroidal field with $\beta =4$,
and is run for 140 orbits at the radius at which the planet will
be inserted.  The relatively strong toroidal field ensures that the
stress will be reasonably large.  Turbulence develops
promptly and the resulting stresses drive accretion.  The component of
greatest importance in the stress tensor is the $R\phi$ component.  For
simplicity, we shall refer to this component as ``the stress.'' The
Reynolds stress needs to be defined carefully, since the concept of
azimuthal velocity fluctuation makes sense only if the mean azimuthal
velocity is known.  But the latter must, in general, be itself
determined from the simulated flow.  We follow \cite{jh00},
and define the Reynolds stress as
\begin{equation}
\langle \rho v_{R} \delta v_{\phi} 
\rangle = {1\over R} 
\left ( \left \langle \frac{(\rho v_R) (\rho Rv_\phi)}{\rho} \right 
\rangle - \frac{\langle \rho v_R \rangle \langle \rho Rv_\phi 
\rangle}{\langle \rho \rangle} \right ).
\end{equation}
The total turbulent stress is a sum of the Reynolds and ($R\phi$) 
Maxwell stresses.

\begin{figure}[htp]
\begin{center}
\psfig{file=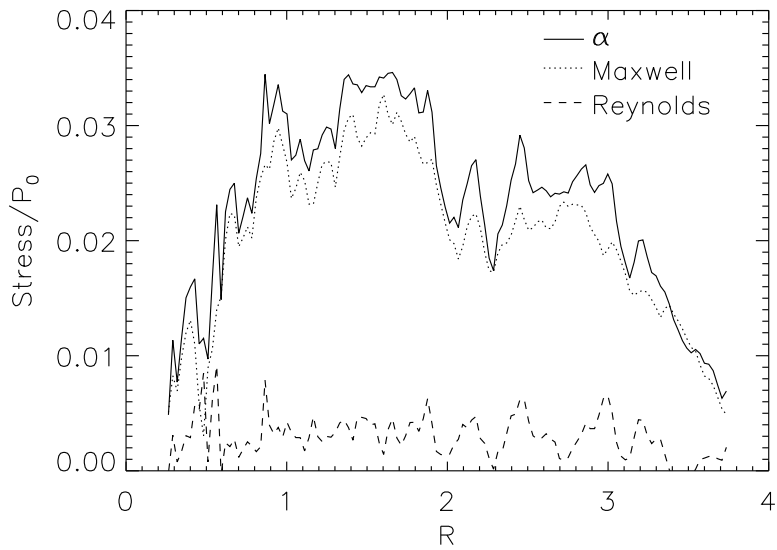,width=4.5in}
\caption{Radial profile (solid line) of the vertically and azimuthally
averaged turbulent stress $\alpha$ at $t=40$ planetary orbits.
The values are scaled in terms of the initial pressure $P_0$.
Plotted separately are the two components of $\alpha$:
the dotted line is the Maxwell stress, $= -B_R B_\phi / 4\pi$, and the
dashed line the Reynolds stress.  Throughout the disk the Maxwell
stress is dominant.
}
\label{ap}
\end{center}
\end{figure}

It is useful to have a dimensionless form for the stress.  This, of course,
is role of the classical $\alpha$ parameter, here defined as
\begin{equation}\label{alpha}
\alpha \equiv {1\over P_{\rm initial}}
\langle \rho v_R \delta v_\phi - {B_R B_\phi\over 4\pi } \rangle,
\end{equation}
the stress divided by the initial pressure.  The $\alpha$ parameter is
particularly useful for comparisons with the prediction of the viscous
gap formation criterion (\ref{eq:viscous}), and with the viscous
hydrodynamic simulations, e.g., \cite{tml96}, \cite{BCLNP99},
\cite{kl99}, and \cite{NPMK00}.  The function $\alpha (R)$ can be
obtained by averaging the stress in accordance with equation
(\ref{avgeq}).  Figure \ref{ap} shows the $\alpha (R)$ profile at
$t=40$ orbits for the baseline toroidal field simulation.  The Maxwell
and Reynolds stresses are also plotted separately.  The Maxwell stress
is clearly dominant, consistent with earlier studies of MRI-driven
accretion disk turbulence \citep{HBW99}.  The diminishing of $\alpha$
at the radial edges is almost certainly a consequence of the initial
and boundary conditions, and the true radial dependence (assuming there
is one) of $\alpha$ cannot be determined with certainty.  One can,
however, extract a reasonable global average, which is $\alpha =
0.02$.  The magnitude scale of $\alpha$, its fluctuation amplitudes,
and its general radial variation are all similar to those found in
other toroidal field cylindrical disk studies \citep{jh01, sp02}.
Translating the globally averaged $\alpha$ into an effective viscosity
and using it in the viscous condition (\ref{eq:viscous}) for gap
formation, one would conclude that only the large mass planet ($q = 5
\times 10^{-3}$) should be able to clear a gap.

\begin{figure}[htp]
\begin{center}
\psfig{file=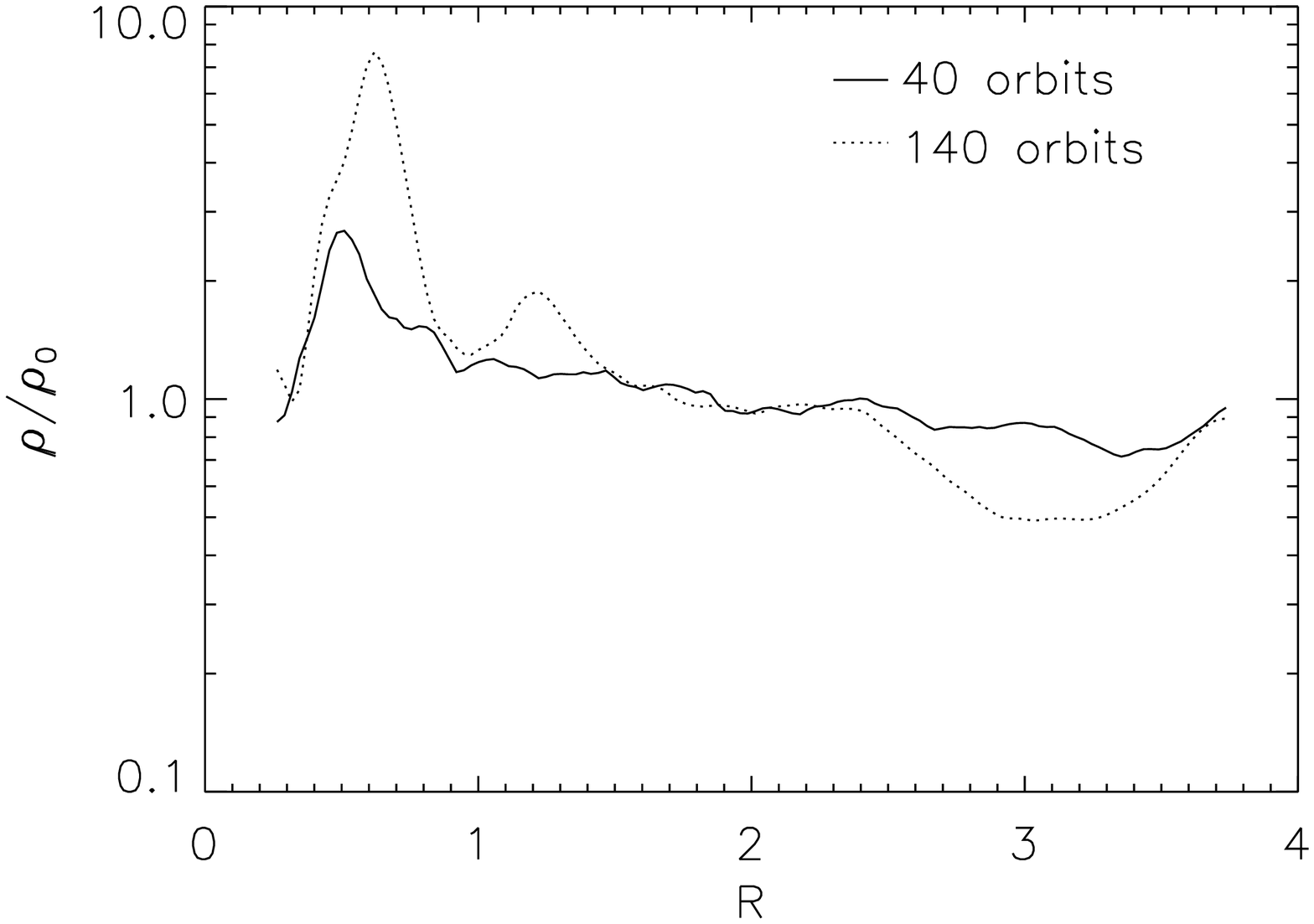,width=4.5in}
\psfig{file=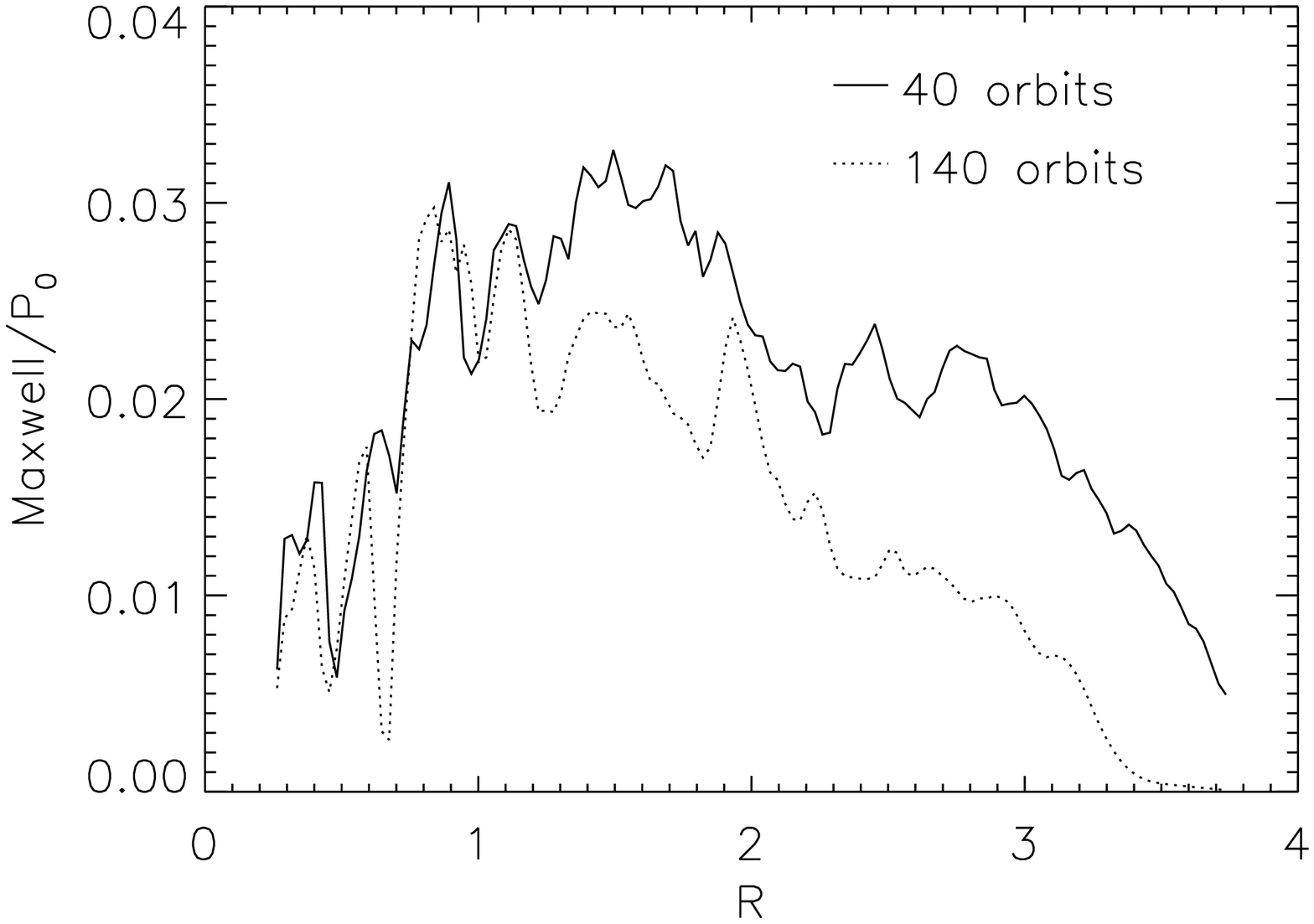,width=4.5in}
\caption{Radial profiles of (a) density and (b) Maxwell stress at
$t=40$ orbits (dotted line) and $t=140$ orbits (solid line)
in the planetless toroidal field simulation MNP.   The change in the
radial density profile is due to the presence of substantial accretion
driven by the turbulent stresses.}
\label{nopds}
\end{center}
\end{figure}

The initial condition for the planet simulations corresponds to orbit
40 of the unforced toroidal field run.  For comparison, the unforced
run is evolved a further 100 orbits.  The top panel of figure
\ref{nopds} shows the radial profile of density normalized by the
initial profile at $t=40$ (dotted line) and 140 orbits (solid line).
The density in the outer disk gradually decreases with time, while the
density in the inner disk increases.  The density at $R=2$, the radius
at which the planet will be inserted, remains nearly constant.

The bottom panel of figure \ref{nopds} shows the radial profile of the
Maxwell stress normalized by the initial pressure.  As the simulation
evolves, the Maxwell stress in the outer disk decreases.  This may be
due in part to the reduction in mass and field energy in this region of
the disk caused by losses through the outer boundary.  As a practical
matter, the time over which the simulation can be evolved is limited to
$~\sim 200$ orbits, by which point significant global evolutionary
changes occur.

While the toroidal field model appears to be best suited to the present
study, it is desirable to investigate other magnetic geometries, such
as a vertical field.  But vertical magnetic fields have proven to be so
vigorously unstable \citep{jh01, sp02},  that large $\alpha$ values and
substantial evolution occur over a relatively short time interval.  In
a cylindrical disk, an initial uniform vertical field can only be lost
through the boundary,
and is otherwise always present to drive the linear ``channel mode''
of the MRI \citep{hb92}.  Vertical fields also tend to open up gaps in
cylindrical disks even without a planet present: the Maxwell stress
increases where the density decreases, and underdense regions become
yet more underdense.

One way around these difficulties is to choose an initial magnetic field
strength that is so small that the fastest growing unstable wavelength is
unresolved.  The field strength is set so that fastest growing wavelength,
$\lambda_{max}$, is set equal to the vertical size of one grid zone,
$\Delta z$, i.e.  \citep{BH98},
\begin{equation}
B_{z}(R) = (15/16)^{1/2}\ \Omega(R)\ (4 \pi \rho)^{1/2} \ 
{\Delta z \over 2\pi}.
\end{equation}
This constraint artificially ``cripples'' the MRI, reducing the linear
growth rate of the resolved wavelengths, and allowing long duration
simulations.  The linear growth rate is proportional to $\sim  v_A/N
\propto \Omega/N$, where $v_A$ is the Alfv\'en speed, and $N$ is the
number of grid zones of the resolved unstable mode.  In the outer
portion of the disk this becomes very low.  Even with this crippled
field, however, there is still some tendency to produce underdense gaps.
The unforced simulation develops radially-alternating high and low density
regions throughout the disk by orbit 140.  The crippled vertical field
model should be regarded not so much as a good model for a disk (it is,
after all, intentionally underresolved), but as a numerical experiment
that provides an alternative turbulent state for comparison with the
toroidal field results.

\subsection{Planets in turbulent disks}

The parameters for the MHD simulations with a planet are listed in
Table \ref{mhdtbl}.  The runs are labeled M for MHD, NP for no planet,
or SP, MP, LP for small, medium and large mass planet, and a Z
indicates a run with the initial vertical field.  Except for the
inclusion of magnetic fields, these planet simulations had the same
parameters as the hydrodynamic tests described in \S3.  The value of
the viscous gap parameter $q_v$ used in Table \ref{mhdtbl} assumes
$\alpha = 0.02$.

\begin{table}[t]
\begin{center}
\caption[MHD Runs]{2.5-D MHD Simulations}
\vskip 12pt
\begin{tabular}{|l|c|c|c|c|c|c|}
\hline
Label & $q$ & ${H}/{R}$ &$q/q_t$ &$q/q_v$& Field Geometry &
$\tau_{\mathrm{gap}}$\\ \hline \hline
MNP & $0$ & 0.05 & -- & --  & toroidal $\beta = 4$ & -- \\
MSP & $2\times 10^{-4}$ & 0.05 &$0.52$&$0.1$& toroidal $\beta = 4$ &
anti-gap \\
MMP & $1\times 10^{-3}$ & 0.05 & $2.6$&$0.5$& toroidal $\beta = 4$ &
215 orbits\\
MLP & $5\times 10^{-3}$ & 0.05 & $13.0$&$2.5$& toroidal $\beta = 4$ & 75
orbits\\
MMPH & $1\times 10^{-3}$ & 0.09 & $0.45$&$0.17$& toroidal $\beta = 12.96$ & 
anti-gap \\
MSPC & $2\times 10^{-4}$ & 0.02 & $8.3$&$0.67$& toroidal $\beta = 0.64$ & 60
orbits \\
MZNP & $0$ & 0.05 & -- & -- & Vertical, $\lambda_{max} = \Delta z$ & -- \\
MZMP & $1\times 10^{-3}$ & 0.05 & $2.6$&$0.5$& Vertical $\lambda_{max} = \Delta z $ &
90 orbits\\
\hline
\end{tabular}
\label{mhdtbl}
\end{center}
\end{table}

\begin{figure}[htb]
\begin{center}
See 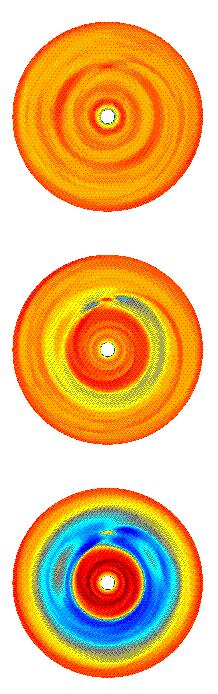
\caption{Log density in toroidal field MHD simulations at $t=100$
orbits.  Plotted are
(a) the small mass planet, $q=2\times 10^{-4}$, (b) the
medium mass planet, $q=1\times 10^{-3}$, and (c) the large mass
planet, $q=5\times 10^{-3}$.
The color map runs from blue to red in $\log(\rho)$ from
$\rho = 0.01$ to 1.0.
}
\label{mhdd}
\end{center}
\end{figure}

\begin{figure}
\begin{center}
\psfig{file=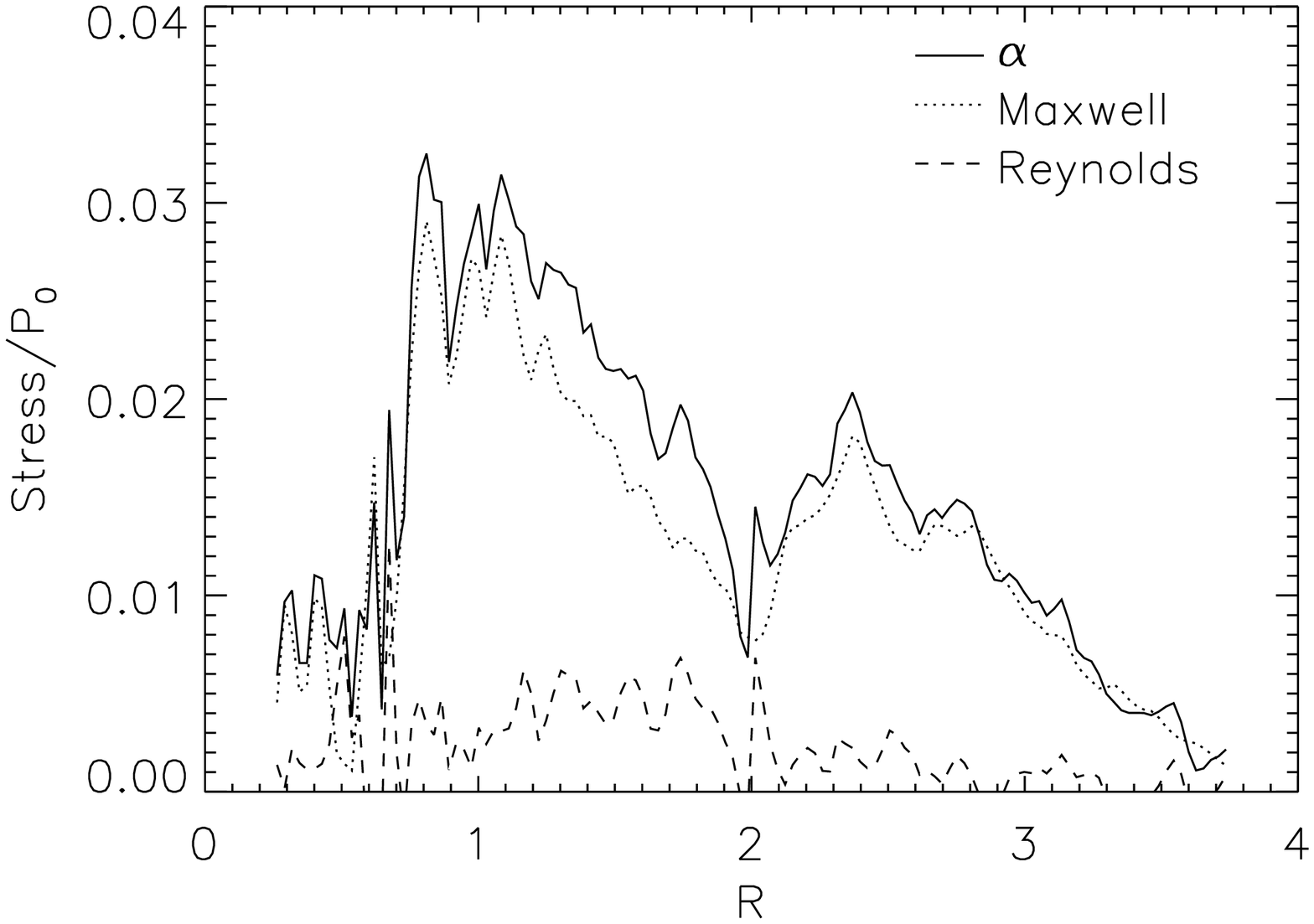,width=3.0in}
\psfig{file=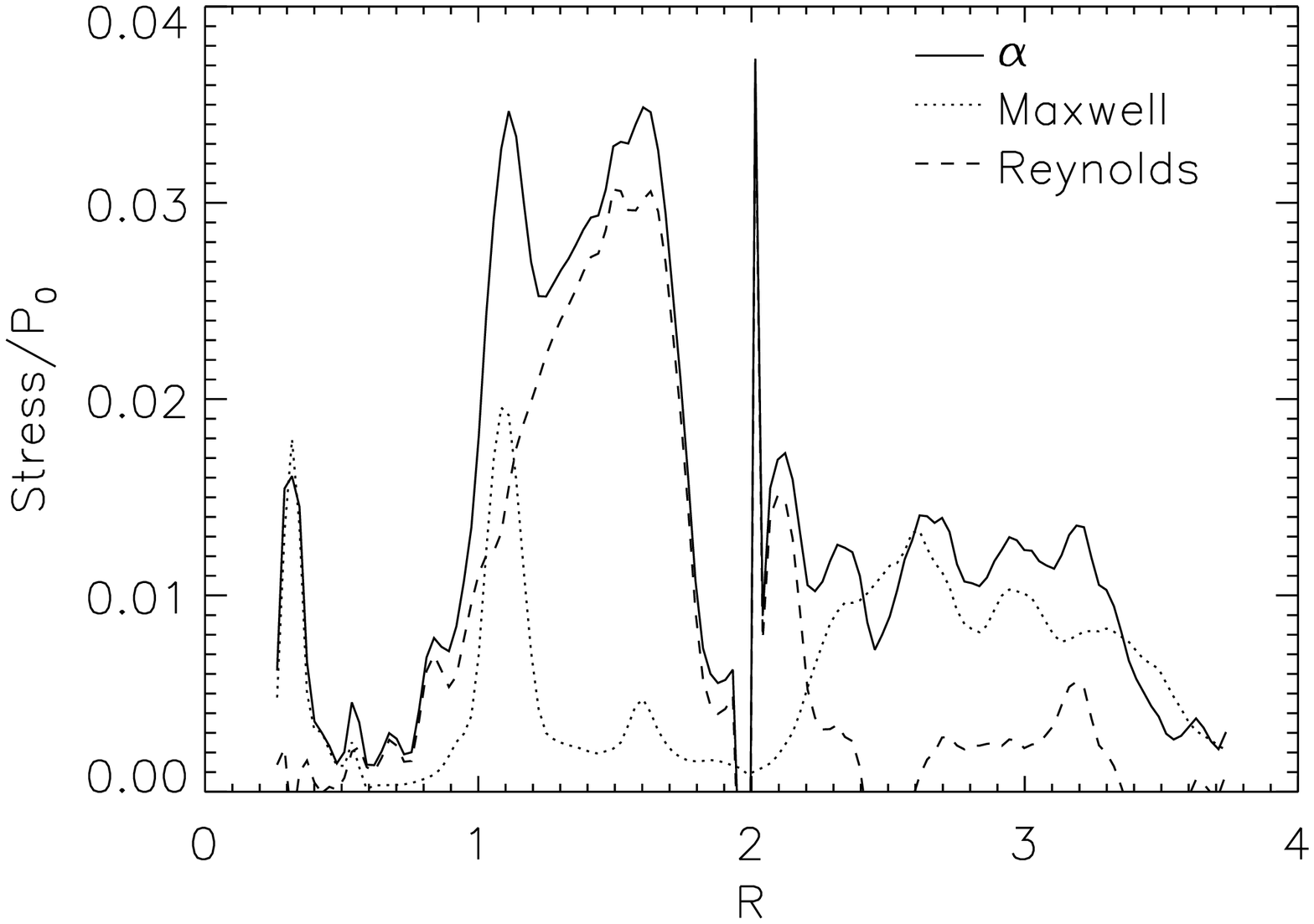,width=3.0in}
\psfig{file=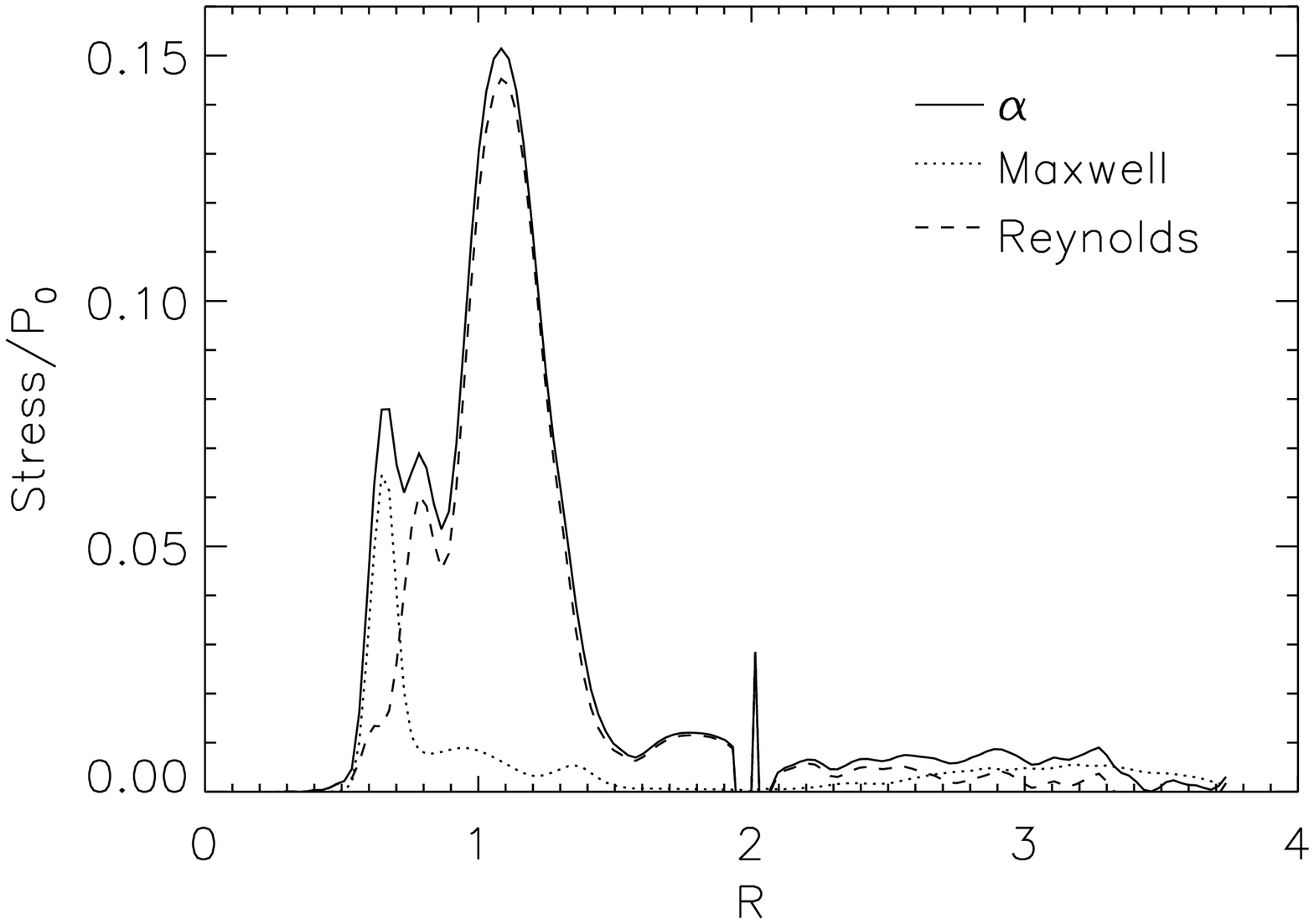,width=3.0in}
\caption{
Radial profile (solid line) of the vertically and azimuthally averaged
turbulent stress $\alpha$ and the two components of $\alpha$:  the
dotted line is the Maxwell stress, $= -B_R B_\phi / 4\pi$, and the
dashed line the Reynolds stress.  The values are normalized to the
initial pressure $P_0$.  (a) Simulation MSP, the small mass planet.
The Maxwell stress remains comparable to the planetless simulation
(fig.~\ref{ap}).  except in the vicinity of the planet's orbit.
(b) MMP, the medium mass planet. The average Maxwell
stress interior to the planet's orbital radius is significantly
reduced
where there is a large Reynolds stress from the spiral wave
driven by the planet.  Overall, total stress $\alpha$ remains
comparable to the planetless simulation.
(c) MLP, the large mass planet simulation.  The average Maxwell
stress interior to the planet's orbital radius is significantly
reduced
and the Reynolds stress from the spiral wave is much larger than the
total stress in the planetless simulation.  The Maxwell stress
increases just where the Reynolds stress dies out.
}
\label{ps}
\end{center}
\end{figure}

Figure \ref{mhdd} shows the density after 100 planetary orbits for runs
MSP, MMP, and MLP, and figure \ref{ps} is the radial distribution of
the Maxwell stress at late time for these simulations.  Figure \ref{ps}
illustrates that the level of turbulent stress is not independent of
the planet; this has consequences.  One such consequence is illustrated
by MSP, a small mass planet that fails to satisfy either the tidal or
the viscous gap opening condition.  In MSP the density {\em increases}
within the Roche radius of the planet, forming  an ``anti-gap.'' The
Maxwell stress has decreased near the planet and gas accretes into this
region faster than it leaves.  This density rise further decreases the
local Alfv\'{e}n velocity of the fluid, reducing the wavelength of the
fastest growing MRI mode, further reducing the stress.  Outside of the
planet's Roche radius the low mass planet produces a weak spiral wave,
but otherwise has a minimal effect on the evolution of the accretion
disk.

\begin{figure}[htp]
\begin{center}
\psfig{file=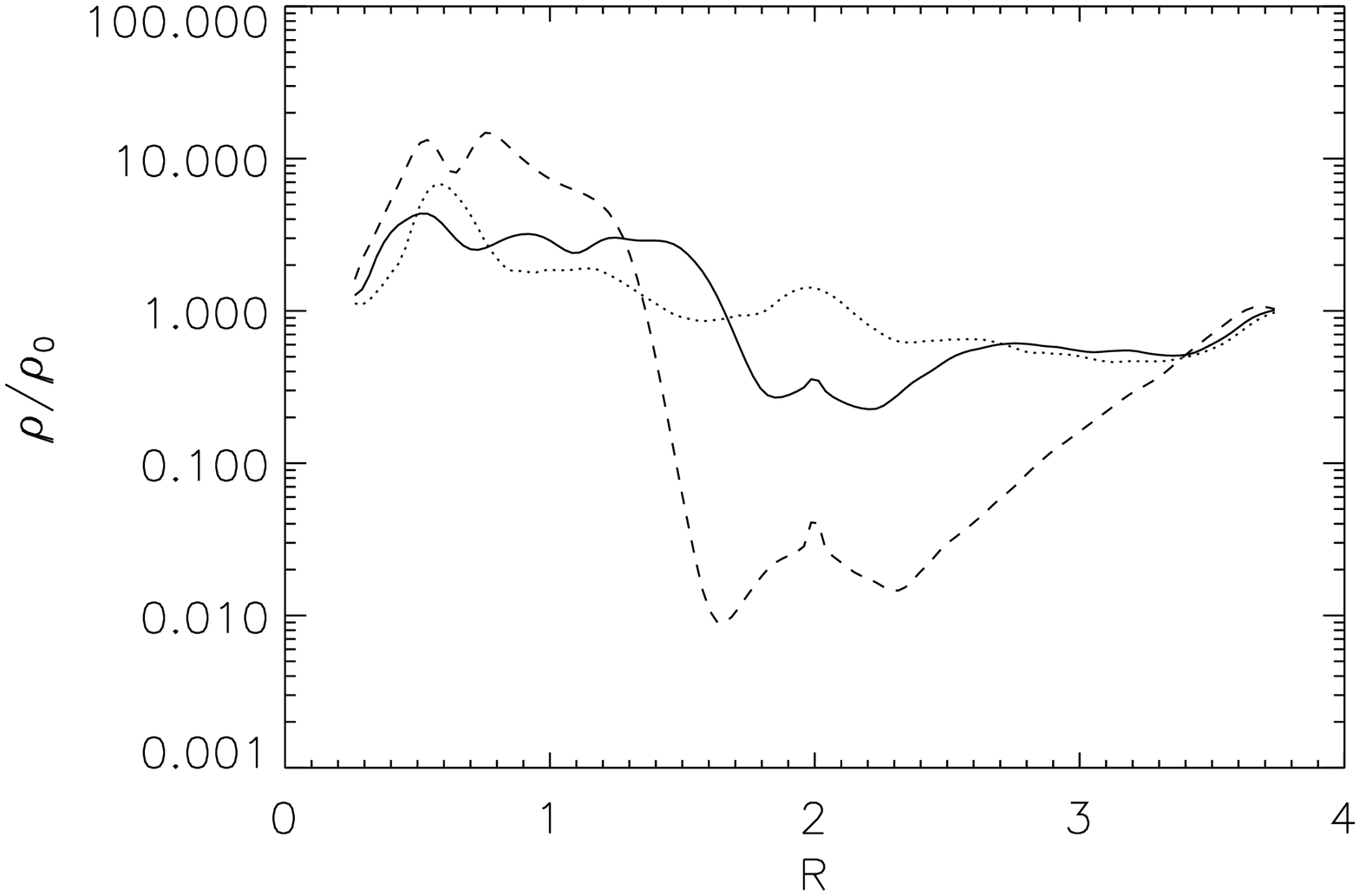,width=4.5in}
\psfig{file=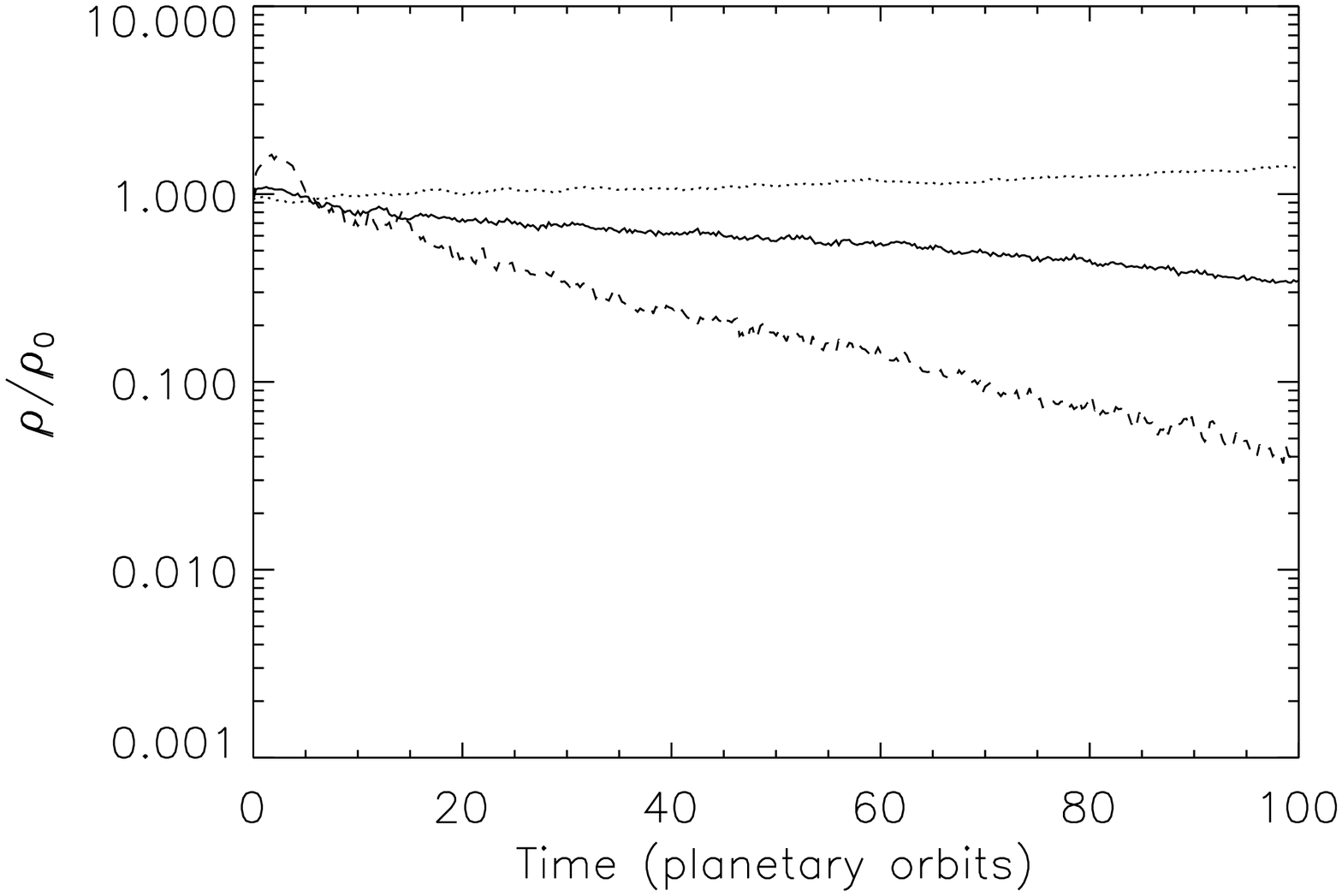,width=4.5in}
\caption{Time evolution of the gap density for toroidal MHD
simulations
MSP (dotted line), MMP (solid line), and MLP (dashed line).
(a) The azimuthally-averaged radial density profiles after
$t=100$ planetary orbits.  (b) The time evolution
of the azimuthally-averaged density at the planet's orbital
radius for each of the three simulations.}
\label{pdt}
\end{center}
\end{figure}

The top panel of figure \ref{pdt} is a comparison of the radial density
profiles for the three embedded planet toroidal MHD simulations,
similar to the top panel of figure \ref{hyddentime}.  The corresponding
plots for each one of the planetary masses are easy to separate by the
depth of the gap produced.  The low mass planet simulation produces a
density increase at the planet's radius, while the other planets' gaps
are shallower than those seen in the hydro runs, and their outer edges
were not as well-defined.  Note that the overall slope in the radial
density profile is consistent with the presence of ongoing accretion.
The bottom panel of figure \ref{pdt} displays the time history of the
density, similar to the bottom panel of figure \ref{hyddentime}.  Once
again, the three different planetary masses can be separated by the
rate at which the density decreases.  The small mass planet shows a
steady increase in density at the planet's orbit.  In the other two
cases, the rate of density decrease is much less than with
hydrodynamics alone; accretion hampers gap formation.

The middle mass planet (run MMP) satisfies the tidal condition but not
the viscous condition.  Exceeding the tidal condition ensures that the
planet will exert a significant influence on the the accretion disk
whether or not it is ultimately able to clear a gap.  As is clear from
figure \ref{mhdd} the planet was able to produce a noticeable density
reduction around its Roche radius.  The planet also creates a strong
spiral wave in the disk.  Most surprisingly, after about 50 planetary
orbits, the magnetic stress inside $R_p$ decreases to very
low levels (see fig.\ \ref{ps}).  Again, not only does the stress in
the disk affect the planet's ability to form a gap, but, at least in
these reduced growth rate simulations, the planet can influence the
turbulent stress within the disk.

The large mass planet (run MLP) easily satisfies both gap criteria,
and its effect on the disk is the most extreme.  A large density gap
develops rapidly, although the formation time $\tau_{\mathrm{gap}}$ is
increased over the purely hydrodynamic model.  A strong global spiral
wave is generated, from the inner to the outer radial boundaries.  And,
as with run MMP, the planet modifies the disk by lowering the Maxwell
stress, here in both the inner and outer disk.

In these three runs the planet was placed into an already turbulent
magnetized disk.  In the next experiment, we approach the problem from
another direction:  we add a magnetic field to a purely hydrodynamical
disk in which a gap has formed.  This new simulation, labeled HMMP,
is initialized by adding a $\beta=4$ toroidal field to the disk from
simulation HMP after 100 planetary orbits.  The result is a disk with
just the same parameters as in simulation MMP.  The ordering of the
appearance of turbulence and the appearance of a planet does not matter.
In MMP, a shallow gap forms even though the mass ratio $q$ does not
exceed the viscous criterion.  Adding a magnetic field to HMP has an
immediate effect, and accretion begins to fill in the much deeper gap
produced without stress.

\begin{figure}[htp]
\begin{center}
\psfig{file=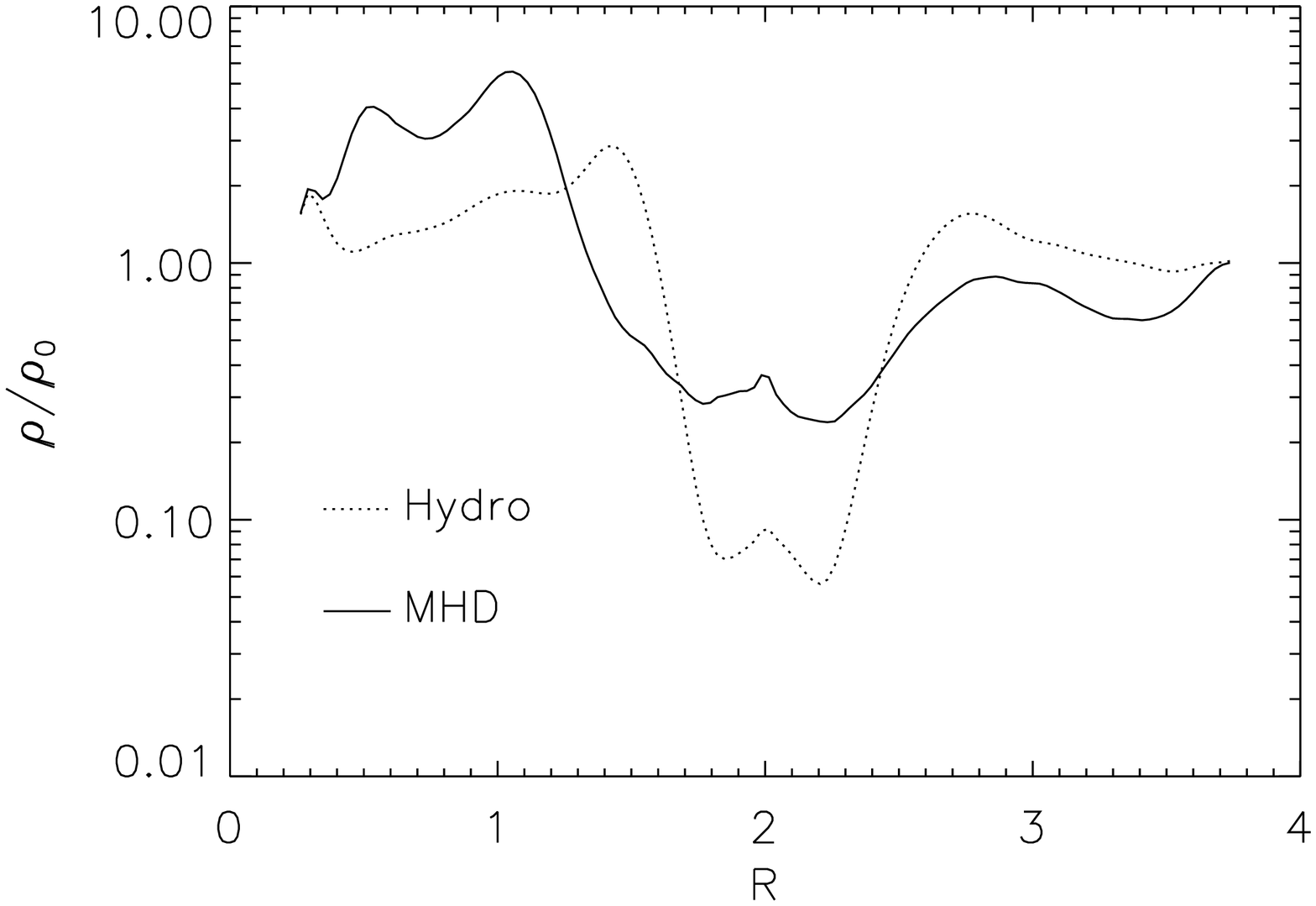,width=4.5in}
\psfig{file=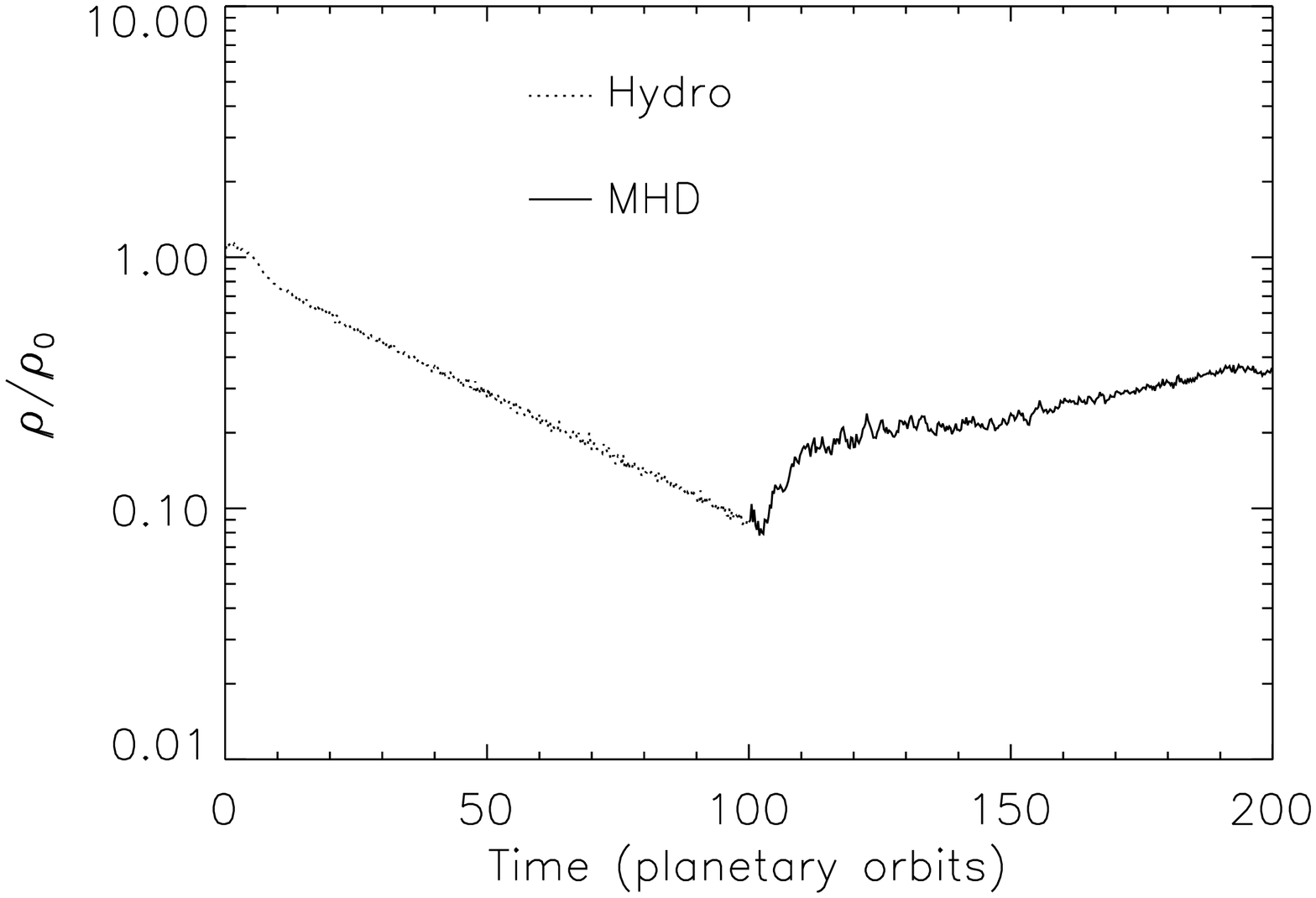,width=4.5in}
\caption
{Density in simulation HMMP.  In this model the first $100$ planetary
orbits are purely hydrodynamical.   A toroidal magnetic
field is then added, MHD turbulence develops, accretion into the gap
begins, and the gap becomes wider and shallower.  (a)
Radial density profile at the end of HMMP compared with the density
at the end of the purely hydrodynamic portion of the run.  (b) Time
history of the density at the orbit of the planet.
}
\label{phmhddentime}
\end{center}
\end{figure}

In figure \ref{phmhddentime}, the top panel shows the radial density
profile at the moment the magnetic field is added to the disk (dotted
line), and after 100 orbits of MHD evolution (solid line).  The bottom
panel depicts the time history of the density at the planet's radius.
The initial hydrodynamic phase exhibits an exponential decrease in
density as the gap deepens.  However, once the toroidal magnetic field
is included and turbulence develops, the density at the planet's radius
begins to increase, returning to about the same level that it had at the
end of run MMP.  At the end of the simulation the gap has become
shallower, wider, and much less well-defined than in the hydrodynamic
simulation.  The turbulent stresses drive accretion across the gap,
increasing the mass in the inner disk at the expense of the outer
disk.  Apparently a planet that satisfies the tidal criterion, but not
the viscous criterion, can form only a partial gap.

We have also examined the effect of varying $H/R$ by altering the
temperature of the underlying disk.  To keep the average $\alpha$ value
comparable to the previous simulations,  the initial magnetic field
strength must also be adjusted.  The ratio of the fastest growing
wavelength to the disk radius, ${\lambda_{max}}/{R}$, is held constant
by varying $\beta$, since
\begin{equation}
\frac{\lambda_{max}}{R} \propto \frac{1}{\sqrt{\beta}}
\left(\frac{H}{R}\right).
\end{equation}

Run MMPH is a hotter disk ($H/R=0.09$) with the medium mass planet.
This planet does not satisfy either the tidal or viscous criterion for gap
formation.  In fact, an anti-gap, or over-density, is produced, just as
in the MSP simulation.  

Run MSPC is a cooler disk with ${H}/{R} = 0.02$, $\beta = 0.64$, and a
small mass planet.  It satisfies the tidal criterion, but not the viscous
criterion.  During the first 60 orbits of evolution, the density drops
inside the planet's Roche radius; after that time the rate of decrease
declines nearly to zero.  The density in the gap levels off at around 8\%
of its original value,  similar to the results of the MMP simulation.
When the tidal criterion is satisfied, the density within the Roche
radius is reduced, but if the viscous condition is not met, the gap
is not completely emptied.

Initially, in run MSPC there is a measurable Maxwell stress throughout
the disk, but once the planet is inserted, the magnetic stress in the
inner disk is reduced.  A comparison between this simulation and MSP
is instructive.  When the planet was not able to produce a density gap
in MSP, the Maxwell stress was largely unaffected.  In the cooler disk,
however, the small mass planet did produce a gap, and the turbulence in
the inner disk was diminished.  This suggests that the threshold for
gap production is closely related to the threshold for affecting the
disk stress.  Evidently, both processes mark the onset of nonlinear
disk-planet interactions.

In run MZMP, a medium mass planet in a vertical field, the disk was
evolved for 100 orbits, allowing a gap to form.  In contrast to run
MMP, the Maxwell stress was sustained inside of the planet's orbit, but
it is significantly reduced outside of the planet's orbit.  Of course
even in run MZNP which does not have a planet, it is clear that at
large radii in the MHD turbulence is rather weak, so this effect cannot
be attributed solely to the presence of the planet.  The continued
presence of Maxwell stresses in the inner portion of the disk, however,
is a significant difference from what is seen in the toroidal field
simulation.

Comparing the density evolution in run HMP with MZMP we find remarkably
good agreement between the two simulations.  Even though the turbulent
stresses are initially present in the MHD simulation, the resulting
density profile and time history are essentially those of the hydro
simulation.  The reason for this is clear.  In the MHD simulation at
late times, the Maxwell stress in the outer disk is drastically
reduced, effectively ending inward accretion.  With no fluid moving
from the outer disk to the inner, nothing flows across the gap.
Consequently, the rate of gap formation and radial profile of the gap
are the same as in the hydro simulation.

\subsection{Effect of planet on stress}

The simulations have provided evidence that not only does the stress
affect the formation of a gap around the planet, but, under some
circumstances, the planet can directly influence the turbulent stress.
In run MSP, the stress was reduced only within the planet's Roche
radius, while in the larger mass planet runs (MMP and MLP), throughout
most of the inner disk the Maxwell stress was significantly reduced at
late times.  This occurs more rapidly with the large mass planet.

The question naturally arises whether the presence of the planet
directly causes the reduction of the MHD stress, or whether changes in
the disk and magnetic field due to the overall evolution have altered
the MRI stability properties of the disk.  To distinguish between these
possibilities we return to the end of run MMP and remove the planet.
The disk, magnetic field, and gap remain, but the spiral wave is
quickly lost once the planet is gone.

\begin{figure}[htp]
\begin{center}
\psfig{file=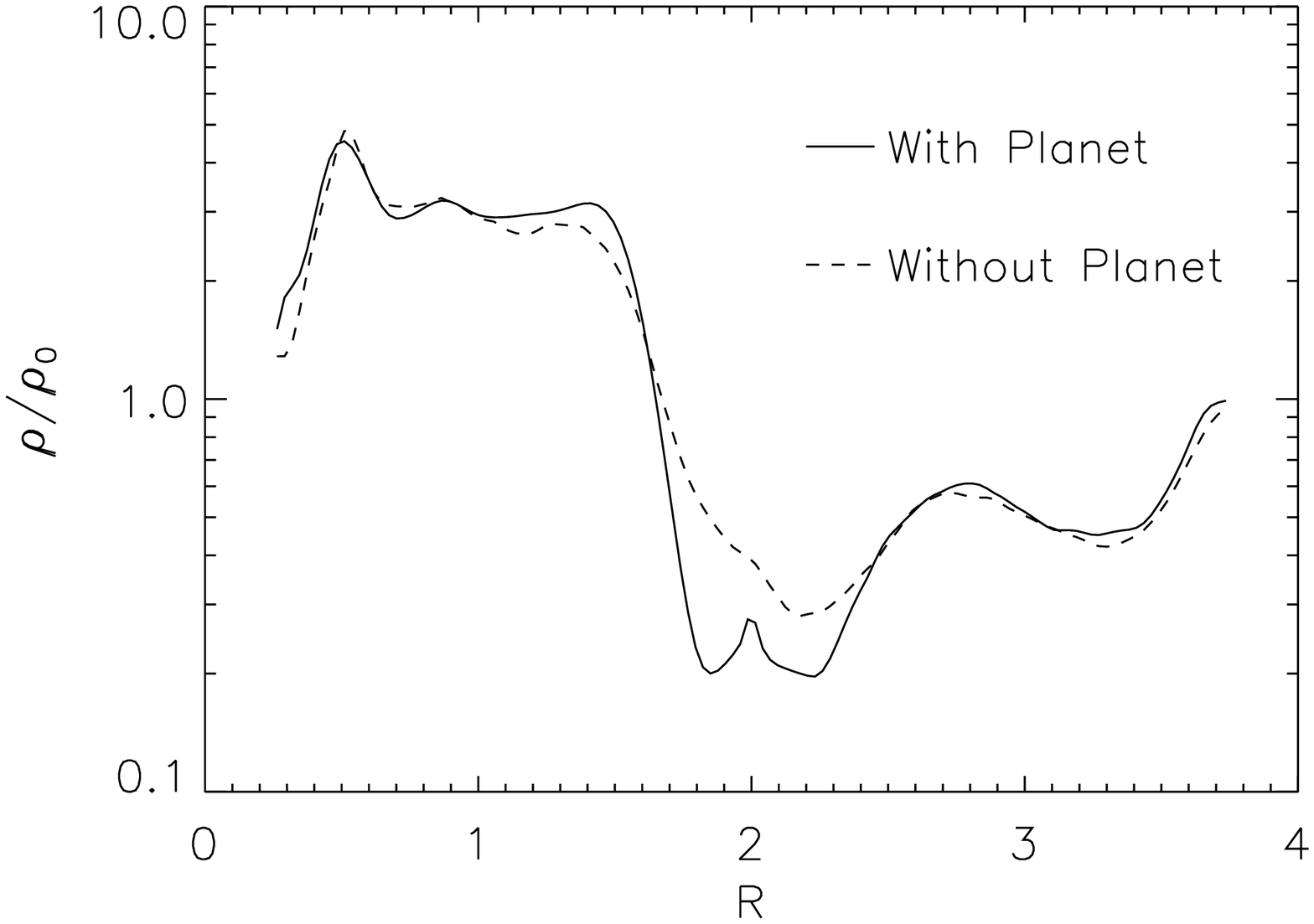,width=4.5in}
\psfig{file=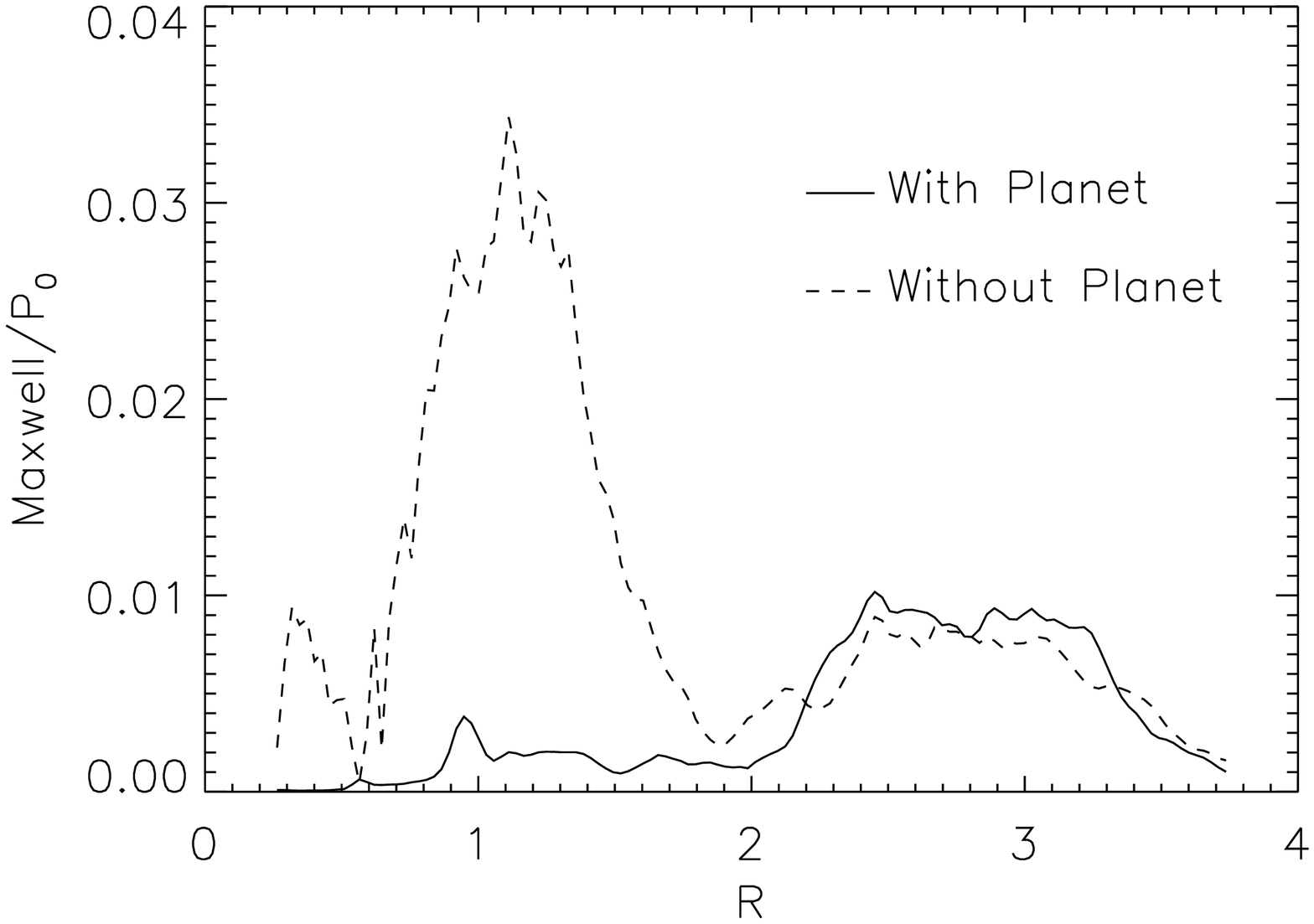,width=4.5in}
\caption{(a) Radial density profile, and (b) Radial profile of the
Maxwell stress ($=-B_R B_\phi / 4\pi$) at the end of run MMP
(solid line),  and 20 orbits of subsequent evolution
after the planet is removed (dashed line).  Once the planet is
removed,
the Maxwell stress interior to the planet's orbit recovers to its
former value
as the Reynolds stress due to the spiral wave dies out.}
\label{rmmpdenstress}
\end{center}
\end{figure}

Figure \ref{rmmpdenstress} shows the before and after radial profiles of
density (top panel) and Maxwell stress (bottom panel).  The solid lines
correspond to the point in time when the planet was removed, and the
dashed lines are from 20 orbits later.  The bottom panel dramatically
shows how the Maxwell stress in the inner disk recovers after the
removal of the planet, increasing by a factor of 10 after 20 orbits.
The density distribution, on the other hand, changes only slightly
over this same time.  The most noticeable change in the radial density
profile is that the gap has become shallower and its edges less defined.
This experiment suggests that interactions between the spiral density
wave and the MHD turbulence affects the level of the Maxwell stress.

\begin{figure}[htp]
\begin{center}
See 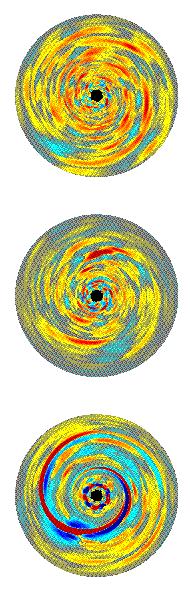
\caption{Contours of the late time, vertically averaged Maxwell stress
from (a) the planetless simulation, (b) the small mass planet
simulation
MSP, and (c) the cold small mass planet simulation MSPC.
Values are scaled by the initial pressure $P_0$.  Negative values are
blues and cyans and run from
$-0.1$ to $0$.  Positive values are yellows and reds and run from $0$
to $0.1$. When the temperature is reduced in
MSPC, the planet's influence on the rest of the disk becomes
stronger and a strong global spiral wave develops.  The presence of
this stron g
wave coincides with the local reduction in the Maxwell stress.
}
\label{max2pi}
\end{center}
\end{figure}

The planet's influence on the disk as a whole is clearly manifest as a
large spiral wave.  Figure \ref{max2pi} displays vertically-averaged
Maxwell stress images from planetless simulation and the two
small-planet MHD simulations, MSP and MSPC.  Without a planet, the
magnetic stress is spread evenly throughout the disk.  In the MSP run,
the planet is not able to form a gap.  The planet produces a spiral
wave that is strong only near the planet.  When the disk is cooled in
the MSPC simulation, the small mass planet is able to form a gap.  As
can be seen from the bottom image, the planet now has a much more
extended influence on the magnetic stress.  The inference is that if
the planet is massive enough to create a gap, it is massive enough to
produce a spiral density wave that interacts with, and lowers, the
turbulent Maxwell stress in the disk interior.

It should be noted that this effect may well be sensitive to the choice
of a toroidal field.  The Maxwell stress in the crippled vertical field
run MZMP was not substantially reduced by the planet, even though it
did form a gap.  This point requires further study.

\section{Conclusions}

In this paper we have examined the formation of gaps by planets in
protostellar disks in the presence of Maxwell and Reynolds stresses
produced by MHD turbulence.  The aims of this first study were to test
the tidal and viscous conditions for gap formation, and to observe the
effects of the planet on an MHD turbulent disk.

We began with a series of inviscid hydrodynamic simulations as
controls.  Under these conditions gap formation is determined by the
tidal criterion: a planet will clear a gap when the local Roche radius is
greater than the pressure scale height.  Our hydrodynamic simulations
agreed with this condition and with the many simulations of such disks
already in the literature.

To establish a turbulent disk, we ran an initial toroidal field
configuration for 40 planetary orbits prior to the insertion of the planet
itself.  The resulting disk had significant Maxwell and Reynolds stresses
that vary with radius, but which gave an average $\alpha$ value of 0.02.
A planet was then added, and the results compared with the tidal and
with the viscous criteria for gap formation.

When the planet was insufficiently massive, or the disk too hot to
satisfy the tidal condition, a gap did not form.  In fact, the density
around the planet increases.  The MHD turbulence drives an overall
accretion flow from the outer to the inner disk, and because the
planet is able to reduce the turbulent stress within its Roche radius,
there is a local buildup in density.

In several simulations, the planet satisfied the tidal criterion, but
not the viscous criterion.  In this case, the density within the Roche
radius is still significantly reduced, but not to zero.  The resulting
gaps are more shallow than those seen in the pure hydro runs, and the
radial density profiles at outer edges of the gap are wider.

When a planet satisfies both the viscous and tidal criteria, a gap
formed, but at a reduced rate compared with the hydrodynamic runs.
Accretion across the gap continues, and the edges of the gap are
somewhat smeared out.  The accretion rate is not large enough to
prevent the density within the gap from being driven to a very low
level.

In a recent paper, \cite{np02b} present a simulation of
a planet placed in an MHD turbulent disk.  Their simulation considered a
$q=5\times 10^3$ mass planet (equivalent to our high mass planet) on a
cylindrical grid with a disk model that has the same radial power law
dependence on $\rho$ and $P$ in the main part of the disk 
as the model as described in \S2.  They use $H/R=0.1$ (hotter than our
models), and an initial sinusoidally-varying vertical field that
produces MHD turbulence with an averaged $\alpha$ value of $0.005$
compared with our models that use a stronger toroidal field with
$\alpha \sim 0.02$.  Nelson \& Papaloizou ran a single simulation with
higher resolution than our models.  The focused on the interaction
of the planet with a disk and a pre-existing gap, whereas our aim is to
consider the gap formation process itself in the presence of MHD
turbulence.

\cite{np02b} found, as we did, that spiral waves produced by
the planet reduced the stresses provided by the MHD turbulence; they
observed this effect mainly near the gap where the waves were strong.
They do not observe evidence for accretion across the gap, whereas some
of our simulations show this.  This suggests that whether or not there
is accretion across the gap depends on such parameters as $q$,
$\alpha$, and $H/R$.

Some results do not differ dramatically from those seen in
viscous hydrodynamic studies.  For example, \cite{kl99} found that the
tidal condition had a greater effect on the width of the gap and the
strength of the spiral waves than variations in viscosity.  By
determining the rate at which gas flows into the gap, the viscosity
affects mainly the accretion rate onto the planet.  \cite{np02b}
compared their model directly with a viscous model and found a number
of differences in detail.  The gap in the turbulent disk model was
deeper than the viscous model with equivalent $\alpha$ value.

Clearly, matters are much more interesting than the viscous models
would predict.  The presence of the planet itself directly influences
the turbulence levels and stress within the disk.  This could not be
seen with a standard $\alpha$ viscosity simulation, for in such a model
the viscosity can influence, but not be influenced by, the planet.  We
found that the small mass planet reduced the Maxwell stress in its
immediate vicinity, allowing matter to accumulate here.  More
significantly, the medium and large mass planets were able to reduce
the turbulence levels in the disk region interior to their orbits.
Tests showed that this was not because the MRI was stabilized by
changes within the disk structure.  Rather, it appears to be due to a
nonlinear interaction between the spiral wave induced by the planet and
the turbulence.  This wave has to exceed a certain strength to have
this effect.  In the main part of the disk the rule is when the planet
is massive enough to open a gap it can also reduce the turbulent
stress.  It should be noted that this effect was observed only in the
disks where the initial field was toroidal, and not in the simulation
with a weak vertical field.

\acknowledgements{We acknowledge support under NSF grant AST-0070979,
and NASA grants NAG5-10655 and NAG5-9266.  Some of the simulations
describe here were carried out on computational platforms supported by
the National Center for Supercomputing Applications and the San Diego
Supercomputer Center. }


\begin{thebibliography}{}


\bibitem[Balbus \& Hawley(1991)]{BH91} Balbus, S. A., \& Hawley, J. F. 
1991, ApJ, 376, 214

\bibitem[Balbus \& Hawley (1998)]{BH98} Balbus, S. A., \& Hawley,
J. F. 1998, Rev Mod Phys, 70, 1

\bibitem[Balbus \& Papaloizou (1999)]{BP99} Balbus, S. A., \& 
Papaloizou, J.~C.~B. 1999, ApJ, 521, 650

\bibitem[Bryden et al. (1999)]{BCLNP99} Bryden G., Chen X., Lin D.,
Nelson R., \& Papaloizou J. 1999, ApJ., 514, 344


\bibitem[Ciecielag, Plewa, \& R\'{o}zyczka (2000)]{CPR00} Ciecielag P.,
Plewa T., \& R\'{o}zyczka M., 2000, in Disks, Planetesimals and
Planets, Garz\'{o}n, C. Eiroa, D. de Winter, and T. J. Mahoney, eds.
(San Francisco: ASP) 219, 45

\bibitem[D'Angelo et al (2002)]{DHK02} D'Angelo G., Henning T., \& Kley
W. 2002, A\&A, 385, 647

\bibitem[Frank, King, \& Raine (1992)]{FKR92} Frank, J., King, A., \&
Raine, D. 1992, Accretion Power in Astrophysics, 2nd ed., (Cambridge:
Cambridge University Press)

\bibitem[Fromang, Terquem, \& Balbus (2002)]{ftb02} Fromang, S.,
Terquem, C., \& Balbus, S. A. 2002, MNRAS, 329, 18

\bibitem[Goldreich \& Tremaine (1980)]{gt80} Goldreich, P., \&
Tremaine, S. 1980, ApJ, 241, 425

\bibitem[Goodman \& Rafikov (2001)]{gr01} Goodman, J., \& Rafikov,
R.~R. 2001, ApJ, 552, 793

\bibitem[Hawley (2000)] {jh00} Hawley, J.~F. 2000, ApJ, 528, 462

\bibitem[Hawley (2001)] {jh01} Hawley, J.~F. 2001, ApJ, 554, 534

\bibitem[Hawley \& Balbus (1992)]{hb92} Hawley, J.~F., \& Balbus, S.~A.
1992, ApJ, 400, 595

\bibitem[Hawley, Balbus, \& Winters (1999)]{HBW99} Hawley, J.~F.,
Balbus, S.~A., \& Winters, W.~F. 1999, ApJ, 518, 394

\bibitem[Hawley \& Stone (1995)]{HS95} Hawley, J. F. \& Stone, J.~M. 
1995, Comp Phys Comm, 89, 127

\bibitem[Kley (1999)]{kl99} Kley, W. 1999, MNRAS, 303, 696

\bibitem[Kley, D'Angelo, \& Henning (2001)]{kl3d} Kley, W., D'Angelo, G., \&
Henning, T. 2001, ApJ, 547, 457

\bibitem[Lin \& Papaloizou (1979)]{lp79} Lin, D.~N.~C., \& Papaloizou,
J.~C.~B.  1979, MNRAS, 188, 191

\bibitem[Lin \& Papaloizou (1993)]{LP93} Lin, D.~N.~C., \& Papaloizou,
J.~C.~B. 1993, in Protostars and Planets III, Levy, E., Lunine,
J. eds. (Tucson: University of Arizona Press), p. 749

\bibitem[Lin \& Papaloizou (1996)]{lp96} Lin, D.~N.~C., \& Papaloizou,
J.~C.~B.  1996, ARAA, 34, 703

\bibitem[Lubow, Seibert, \& Artymowicz (1999)]{LSA99} Lubow S., Seibert
M., \& Artymowicz P. 1999, ApJ., 526, 1001

\bibitem[Nelson, \& Papaloizou (2002a)]{np02a} Nelson, R., \&
Papaloizou, J.~C.~B. 2002a, MNRAS, in press

\bibitem[Nelson, \& Papaloizou (2002b)]{np02b} Nelson, R., \&
Papaloizou, J.~C.~B. 2002b, MNRAS, in press

\bibitem[Nelson et al. (2000)]{NPMK00} Nelson R., Papaloizou, J., Masset,
F., \& Kley, W. 2000, MNRAS, 318, 18


\bibitem[Shakura \& Sunyaev (1973)]{SS73} Shakura, N. I., \& Sunyaev,
R. A. 1973, A\&A, 24, 337

\bibitem[Steinacker \& Papaloizou (2002)]{sp02} Steinacker, A., \&
Papaloizou, J.~C.~B. 2002, ApJ, 571, 413

\bibitem[Stone \& Balbus (1996)]{sb96} Stone, J. M., \& Balbus, S.
A. 1996, ApJ, 464, 364

\bibitem[Stone \& Norman (1992a)]{SN92a} Stone, J. M., \& Norman, M. L. 
1992a, ApJS, 80, 753

\bibitem[Stone \& Norman (1992b)]{SN92b} Stone, J. M., \& Norman, M. L.
1992b, ApJS, 80, 791 

\bibitem[Takeuchi, Miyama, \& Lin (1996)]{tml96} Takeuchi, T., Miyama,
S. M., \& Lin, D. N. C. 1996, ApJ, 460, 832

\bibitem[Tanaka, Takeuchi, \& Ward (2002)]{ttw02} Tanaka, H.,
Takeuchi, T., \& Ward, W.~R. 2002, ApJ, 565, 1257

\bibitem[Ward (1986)]{w86} Ward, W.~R. 1986, Icarus, 67, 164

\bibitem[Ward (1997)]{wd97} Ward, W.~R. 1997, Icarus, 126, 261

\end{thebibliography}
\end{document}